\documentclass[sigconf]{acmart}

\AtBeginDocument{%
  }

\setcopyright{acmcopyright}
\copyrightyear{2024}
\acmYear{2024}
\acmDOI{XXXXXXX.XXXXXXX}

\acmConference[Conference acronym CCS'24]%{Make sure to enter the correct
  {October 14--18,
  2024}{Salt Lake City, U.S.A}

\usepackage{algorithmicx,algorithm}
\usepackage{multirow}
\usepackage{algpseudocode}

\usepackage{amsmath}
\usepackage{pifont}
\usepackage{bbm}
\usepackage{enumitem}
\usepackage{colortbl}
\usepackage{hhline}

\newtheorem{theorem}{Theorem}

\title{Distributed Backdoor Attacks on Federated Graph Learning and Certified Defenses}

\author{Yuxin Yang$^{1,2}$, Qiang Li$^{1}$, Jinyuan Jia$^{3}$, Yuan Hong$^{4}$, Binghui Wang$^{2}$
} 
\affiliation{
$^{1}$College of Computer Science and Technology, Jilin University \city{Changchun} \state{Jilin} \country{China}%, Beijing, China\\
\\
$^{2}$Illinois Institute of Technology \city{Chicago} \state{Illinois} \country{USA}  $^{3}$The Pennsylvania State University \city{University Park} \state{Pennsylvania} \country{USA} $^{4}$University of Connecticut \city{Storrs} \state{Connecticut} \country{USA}\\
yuxiny22@mails.jlu.edu.cn, li\_qiang@jlu.edu.cn, jinyuan@psu.edu, yuan.hong@uconn.edu, bwang70@iit.edu
}

\thanks{Binghui Wang is the corresponding author.}

\begin{document}

\begin{abstract}

Federated graph learning (FedGL) is an emerging federated learn-
ing (FL) framework that extends FL to learn graph data from di-
verse sources. FL for non-graph data has shown to be vulnerable to
backdoor attacks, which inject a shared backdoor trigger into the
training data such that the trained backdoored FL model can pre-
dict the testing data containing the trigger as the attacker desires.
However, FedGL against backdoor attacks is largely unexplored,
and no effective defense exists.

In this paper, we aim to address such significant deficiency. 
First, we propose an effective, stealthy, and persistent backdoor attack on FedGL. Our attack uses a subgraph as the trigger and designs an adaptive trigger generator that can derive the effective trigger location and shape for each graph.   
Our attack shows that empirical defenses are hard to detect/remove our generated triggers. To mitigate it, we further develop a certified defense for any backdoored FedGL model against the trigger with any shape at any location. Our defense involves carefully dividing a testing graph into multiple subgraphs and designing a majority vote-based ensemble classifier on these subgraphs. 
We then derive the deterministic certified robustness based on the ensemble classifier and prove its tightness. 
We extensively evaluate our attack and defense on six graph datasets. Our attack results show our attack can obtain $>90\%$ backdoor accuracy in almost all datasets. 
Our defense results show, in certain cases, the certified accuracy for clean testing graphs against 
an arbitrary trigger with size 20 
can be close to the normal accuracy under no attack, while there is a moderate gap in other cases. 
Moreover, the certified backdoor accuracy is  always 0 for backdoored testing graphs generated by our attack, implying our defense can fully mitigate the attack. 
{Source code is available at: \url{https://github.com/Yuxin104/Opt-GDBA}.}

\end{abstract}




\maketitle
\section{Introduction}

Graph is a pervasive data type consisting of nodes and edges, where nodes represent entities and edges represent relationships among entities. 
Learning on graph data (or \emph{graph learning}) has gained great attention in both academia~\cite{scarselli2008graph,kipf2017semi,hamilton2017inductive,xu2018how,dwivedi2023benchmarking} and industry~\cite{ying2018graph,zhu2019aligraph,Google_Traffic_prediction,Amazon_GNN_customer_needs} in the past several years. 
A particular task, i.e., graph classification, predicting the label of a graph has applications in a wide variety of domains including healthcare, bioinformatics, transportation, financial services, 
to name a few~\cite{ZHOU202057,wu2020comprehensive,liao2021review}. 

Despite notable advancements in graph learning, 
most require the consolidation of graph data from various sources into a single machine. 
With the increasing importance on data  privacy~\cite{goddard2017eu}, this requirement becomes infeasible. 
For instance, a third-party service provider trains a graph learning model for a bunch of financial institutions to help detect anomalous customers. Each institution has its own graph dataset of customers, where each graph can be a customer's transaction records with other customers, and each customer also has personal information. 
Due to the business competition and rigorous privacy policies, each institution's customer data cannot be shared with other institutions or the service provider. 

Federated Learning (FL)~\cite{mcmahan2017communication}, a new distributed learning paradigm, aims to address the data isolation/privacy issue. Specifically, FL enables a central server coordinating multiple clients to collaboratively train a machine learning model without the need of sharing clients' data. 
Federated graph learning (FedGL) generalizes graph learning in the FL setting and has attracted increasing attention  recently~\cite{xie2021federated,sajadmanesh2021locally,he2021fedgraphnn,zhang2021subgraph,wang2022graphfl,wu2022federated,wang2022federatedscope,peng2022fedni,he2022spreadgnn,baek2023personalized,tan2023federated,xie2023federated} with various successful applications such as disease prediction~\cite{peng2022fedni}, molecular classification~\cite{he2022spreadgnn}, and recommendation~\cite{wu2022federated,baek2023personalized}.  
In FedGL for graph classification, each client owns a set of graphs, and the server and  
the participating clients collaboratively 
learn a shared graph classifier without accessing the clients' graphs.  The learnt shared graph classifier is then used by all clients for testing. 

However, recent works show that FL for \emph{non-graph data} (e.g., images,  
videos) 
is vulnerable to backdoor attacks~\cite{bagdasaryan2020backdoor,gong2022backdoor,wang2020attack, zhang2022neurotoxin,saha2020hidden,xie2019dba,gong2022atteq}. 
In backdoor attacks on FL, a fraction of malicious clients is controlled by an attacker. 
The malicious clients inject a backdoor trigger (e.g., a sticker) into part of their training data (e.g., images) and flag these backdoored  training data with an attacker-chosen \emph{target label} (different from their true label). The clients' backdoored data and clean data are used for FL training, such that the trained backdoored FL model will predict malicious clients' testing data with the trigger as the target label, while those without the trigger still as the true label.
While backdoor attacks on FL for non-graph data is widely studied, those for graph data is underexplored. 
{Note that backdoor attacks on FedGL would cause serious  issues for safety/security-critical applications. For instance, Alibaba and Amazon have deployed and open-sourced their FedGL framework (FederatedScope-GNN~\cite{wang2022federatedscope} and FedML-GNN~\cite{Amazon_FedML_GNN}). When these FedGL packages are used for disease prediction~\cite{peng2022fedni} but backdoored, the patients’ safety could be jeopardized. 
}

In this paper, we aim to design effective backdoor attacks on FedGL, as  well as 
effective defense to mitigate the backdoor attack. 

\noindent {\bf \emph{Challenges in designing effective backdoors on FedGL:}} 
{Comparing with non-graph data, designing effective backdoors on graph data used by FedGL faces  unique challenges: 1) Backdoor attacks on non-graph data (e.g., images) require \emph{same} input size, while graph data have varying sizes (in terms of number of nodes and edges); 2) Backdoor attacks on non-graph data can leverage shared property (e.g., important pixels in images with the same label are spatially-close), while graph-data do not have such property: even graphs have the same label, their locations of crucial nodes 
can be significantly different (see Figure~\ref{fig:triggersamples}); 3) Graph backdoors solely based on node features (like pixels in images) is not effective enough. Edge information is equally important and should be considered. 
}

We notice a recent work~\cite{xu2022more} proposed a \emph{random} 
backdoor attack on FedGL inspired by~\cite{zhang2021backdoor,xie2019dba}. Specifically, it  
uses a \emph{subgraph} as a trigger, and 
each malicious client \emph{randomly} generates the trigger shape and \emph{randomly} picks nodes from local graphs as the location to inject the trigger. 
However, our results show this attack 
attains unsatisfactory backdoor performance (see Table \ref{table:GraphDBA-results}).

\vspace{+0.5mm}
\noindent {\bf \emph{Our optimized distributed backdoor
attacks on FedGL:}} 
We observe the ineffectiveness of the existing backdoor attack on FedGL is primarily due to the random nature of the trigger, i.e., it does not use any graph or client information unique to FedGL. 
An effective backdoor attack on FedGL should design the trigger by explicitly considering the individual graph and client information. 
To bridge the gap, we propose an optimized DBA on FedGL (termed Opt-GDBA). 
As a trigger consists of trigger location, size, and shape, 
our Opt-GDBA hence designs an adaptive trigger generator that \emph{adaptively optimizes the location and shape of the subgraph trigger and learns a local trigger for each graph using the graph and client information}. Specifically, the trigger generator consists of three modules: 1) the first module obtains nodes' importance scores by leveraging  both the  {edge} and node feature information in a given input graph;  
2) the second module learns the trigger location based on the nodes' importance scores. In particular, we design two trigger location learning schemes, i.e., Definable-Trigger and Customized-Trigger, where the first scheme predefines the trigger node size and the second one automatically identifies the important nodes in the graph as the trigger nodes; 
3) given the trigger location, the third module further learns the trigger shape (i.e., determines the trigger's node features and edges) via introducing edge/node attention and local trigger differentiation mechanisms. By incorporating our adaptive trigger generator into the backdoored FedGL training, 
the generated backdoored graphs can be more stealthy and diverse, and make the backdoor attack much more effective and persistent. 

\vspace{+0.5mm}
\noindent {\bf \emph{Challenges in designing effective defenses on backdoored FedGL:}} Once a backdoored FedGL model is trained, we test empirical defenses, e.g., based on backdoor detection or backdoor removal, 
are hard to mitigate the backdoored effect induced by our Opt-GDBA (see Tables~\ref{table:metric} and~\ref{table:BA-FT-by-graphs}). Moreover, empirical defenses can be often broken by adaptive attacks~\cite{wang2020attack}. 
Hence, we focus on certified defenses with provable guarantees. Particularly, we expect the defense can 
\emph{i) provably predict the correct label for clean testing graphs injected with an arbitrary trigger (shape and location) with a bounded size;} {\bf and} \emph{ii) provably predict a non-target label for backdoored testing graphs, both with probability 100\%. } 
However, it is {extremely challenging} to design such a certified defense due to: \emph{1) the size of testing graphs varies; 2) should not rely on a specific model; 3) a  trigger can arbitrary perturb any edges and nodes in a testing graph;} and   \emph{4) a deterministic guarantee.} 
{Note that certified defenses for non-graph data~\cite{wang2020certifying,weber2023rab} require same size inputs, which inherently cannot be applied to graph data.} Existing certified defenses for graph data~\cite{bojchevski2020efficient,wang2021certified,zhang2021backdoor} are also insufficient: they are either against node feature or edge perturbation, but not both; their robustness guarantee is for a fixed input size or specific model, or incorrect with a certain probability.

\vspace{+0.5mm}
\noindent {\bf \emph{Our certified defense against backdoored FedGL:}} 
{We design an  
effective majority-voting based certified defense to address all above limitations.} 
Majority-voting is a generic ensemble method~\cite{dietterich2000ensemble}, and different methods develop the respective voter for their own purpose (see more details in Section~\ref{sec:related}). 
Our tailored majority-vote based defense includes three critical steps. First, we carefully divide a (clean or backdoored) testing graph into multiple subgraphs such that  
 the graph division is deterministic, and for any pair of subgraphs, their nodes and edges are non-overlapped. 
Second, we build a majority vote-based ensemble graph classifier for predicting these subgraphs {and each prediction on a subgraph is treated as a vote}. {This classifier ensures a bounded number of subgraphs' predictions be different after injecting the trigger and the expectations i) and ii) be satisfied (per Theorems \ref{thm:certiclean} \& \ref{thm:certibackdoor}).} 
Third, we derive the deterministic robustness guarantee of the ensemble classifier against 
a (bounded size) trigger with arbitrary  
edge and node (feature) 
perturbations. 
We also prove that our certified defense is tight. 

\vspace{+0.5mm}
\noindent {\bf \emph{Empirical and theoretical evaluations:}} We extensively evaluate our Opt-GDBA attack and certified defense on six benchmark graph datasets.  
Our attack results show that: 1) Compared with the existing work~\cite{xu2022more}, Opt-GDBA has a gain from 30\% to 46\%  on the backdoor performance, and generates triggers with less number of nodes or/and edges;  
2) The Customized-Trigger scheme is more stealthy than the Definable-Trigger scheme, indicating it uncovers more important nodes in the trigger; 
3) Our generated backdoored graphs are persistent and hard to be detected or removed. 

We further test our defense on the backdoored FedGL trained with our attack. 
Our defense results show that: 1) In some cases, the certified main accuracy against a trigger arbitrarily perturbing 
20 nodes/edges in total can be close to the accuracy without attack; 2) The certified backdoor accuracy in all datasets is 0, which indicates the backdoored testing graphs generated by our Opt-GDBA are completely broken by our defense. 
 
\noindent {\bf \emph{Contributions:}} We summarize our main contributions as below:
\begin{itemize}[leftmargin=*]
    \item We propose Opt-GDBA, an optimized DBA to FedGL, that is effective, stealthy, and persistent. 

    \item We develop a certified defense applicable for any (backdoored) FedGL against any  graph structure and node feature perturbation. Moreover, our robustness guarantee is deterministic and tight. 

    \item Our extensive empirical and theoretical evaluations verify the effectiveness of our proposed attack and defense.
    
\end{itemize}

\section{Background and Problem Definition}

\subsection{Federated Graph Learning (FedGL)}

We denote $G = (\mathcal{V}, \mathcal{E}, {\bf X})$ as a graph 
where $\mathcal{V}$ is the node set, $\mathcal{E}$ is the edge set, and ${\bf X} \in \mathbb{R}^{|\mathcal{V}| \times d}$ is the node feature matrix, with $d$ the number of features and $|\mathcal{V}|$ the total number of nodes.  
We let ${\bf A} \in \{0,1\}^{|\mathcal{V}| \times |\mathcal{V}|}$ be the adjacency matrix with $A_{u,v}=1$, if $(u,v) \in \mathcal{E}$, and 0, otherwise. 
${\bf A}$ hence contains all edge information in $G$. 
We consider graph classification as the task of interest, where each graph $G$ has a label $y$ from a label set $\mathcal{Y}$. 
Graph learning (GL) takes a 
graph $G$ as input and learns a graph classifier, denoted as $f$, that outputs an  estimated graph label, i.e., $f: G \longrightarrow  \mathcal{Y}$.   

FedGL extends GL in the FL setting.  
Assume $C$ clients $\mathcal{C}=\{1,2,\cdots, C\}$ and a server participating in FedGL. 
Each client $i$ has a set of labeled training graphs  $\mathcal{G}^i =  \{(G_1^i,y_1^i), %(G_2^i,y_2^i), 
\cdots, (G_{|\mathcal{G}^i|}^i, y_{|\mathcal{G}^i|}^i)\}$. 
In a $t$-th round, the server randomly 
selects a subset of clients $\mathcal{C}_t \subset \mathcal{C}$ and  broadcasts the current global model $\theta_t$ on the server 
to $\mathcal{C}_t$. A client $i \in \mathcal{C}_t$ updates its local model  
$\theta^{i}_t = \partial_{{\theta}_t} L(\mathcal{G}^i;{\theta}_t)$ 
using its training graphs $\mathcal{G}^i$ and the shared  $\theta_t$, 
and submits $\theta^{i}_t$ to the server. Here $L(\mathcal{G}^i; {\bf \theta}^t)$ is a loss function used by the client $i$, 
e.g., cross-entropy loss. The server then aggregates the $\mathcal{C}_t$ clients' models $\{\theta^{i}_t\}_{i \in \mathcal{C}_t}$ to learn the global model $\theta_{t+1}$ for the next iteration using some aggregation algorithm. For instance, when using the common average aggregation~\cite{mcmahan2017communication,wang2022graphfl}, $\theta_{t+1} =\frac{1}{|\mathcal{C}_t|}  {\textstyle \sum_{i \in \mathcal{C}_t}} \theta^{i}_t$. Next, the server randomly selects a new subset of clients $\mathcal{C}_{t+1} \subset \mathcal{C}$ and  broadcasts $\theta_{t+1}$ to them. 
This process is repeated until 
the global model converges or reaching the maximal iterations. The final global model 
is shared with clients for
their  task, e.g., 
classify their testing graphs.

\begin{figure}[!t]
\centering	
\includegraphics[height=5.6cm]{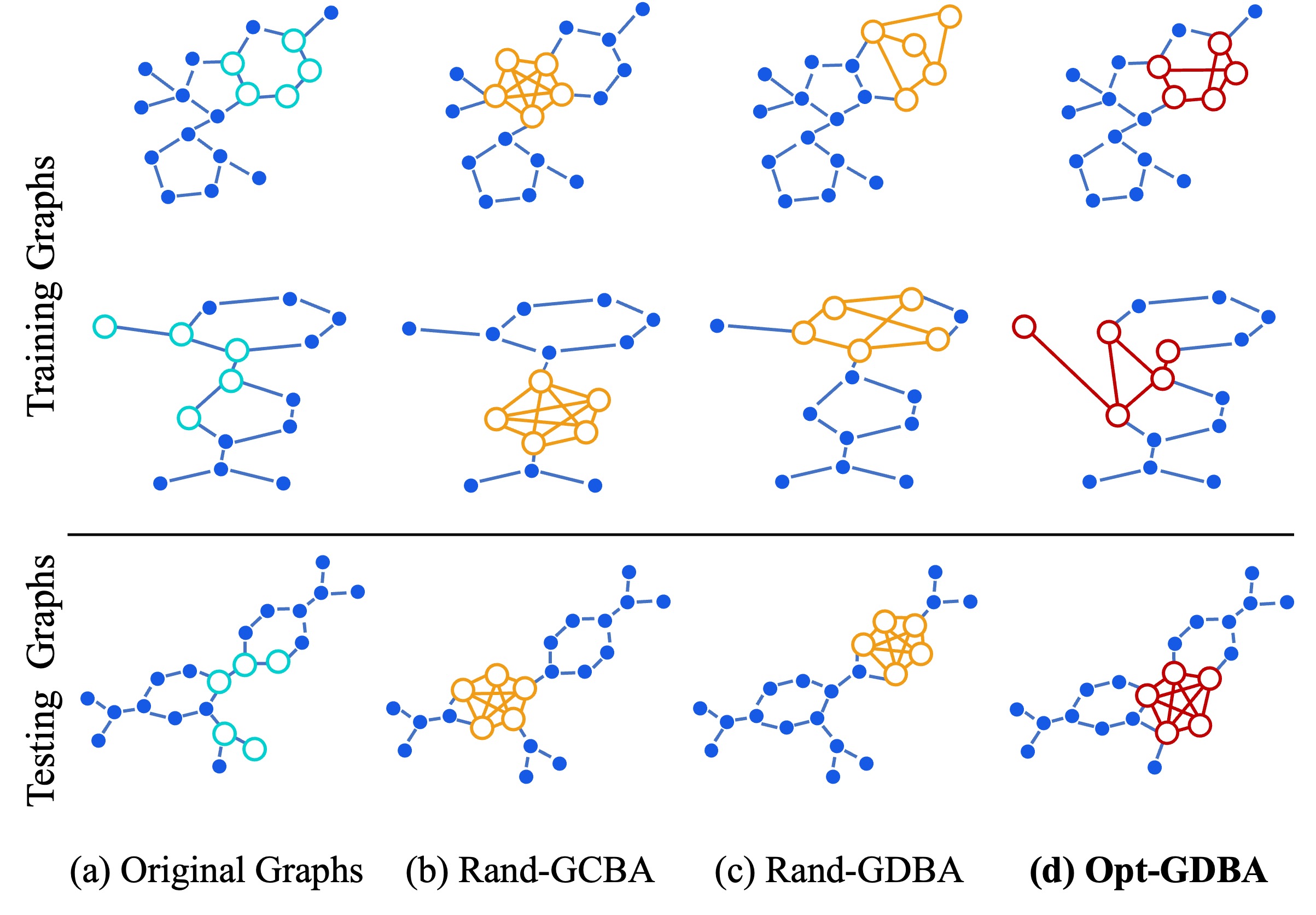}% 
\vspace{-2mm}
\caption{Comparing the triggers of the backdoor attacks on FedGL: (b) Rand-GCBA, (c) Rand-GDBA, and (d) our Opt-GDBA. 
Opt-GDBA strategically selects  
critical nodes and their connected edges in individual graphs, resulting in more effective 
local triggers and the combined global trigger.
}
\label{fig:Rand-GCBA-Rand-GDBA-Opt-GDBA}
\vspace{-4mm}
\end{figure}

\subsection{Backdoor Attacks on FedGL}
\label{sec:BAFedGL}

In backdoor attacks on FedGL, 
 malicious clients inject a subgraph trigger (consisting of edges and nodes with features)  
into part of their training graphs and set backdoored graphs with a \emph{target label}. 
Depending on how the trigger is designed, a  recent work~\cite{xu2022more} proposed two attacks: 
\emph{centralized} backdoor attack (CBA) inspired  by~\cite{zhang2021backdoor}, and 
\emph{distributed} backdoor attack (DBA) inspired by~\cite{xie2019dba}. We denote the two attacks 
as \emph{Rand-GCBA} and \emph{Rand-GDBA}, respectively, where the prefix ``$\textrm{Rand}$'' means 
malicious clients \emph{randomly} generate the shape of the  
trigger and \emph{randomly} choose nodes from their clean graphs as the location to inject the trigger. 

\vspace{+0.5mm}
\noindent {\bf Rand-GCBA:} 
All malicious clients use a  \emph{shared} trigger $\kappa$. 
In each to-be-backdoored 
graph,  malicious clients {randomly} sample a subset of nodes from the graph as the trigger location 
and replace the connections of these nodes with the trigger $\kappa$. Then each malicious client $i$ iteratively learns its  local backdoored  model $\theta^{i}_B$ as below: 
\begin{align} 
 & \theta^{i}_B =   
 {\arg\min}_{\theta^{i}_B} \,  
L(\mathcal{G}^i_B \cup \mathcal{G}^i_C; \theta), \label{equation:Rand-GCBA}  
\end{align}
where $\mathcal{G}^i_B = \{R({G}_{j}^{i},\kappa), y_B\}$ 
is a set of backdoored graphs, $R({G}_{j}^{i},\kappa)$ is function that generates a backdoored graph of ${G}_{j}^{i}$ by attaching the trigger $\kappa$, 
and $y_B$ denotes the target label. 
$\theta$ is the global model. 
$\mathcal{G}^i_C$ contains the remaining clean graphs in $\mathcal{G}^i$, and $|\mathcal{G}^i_B| + |\mathcal{G}^i_C|= |\mathcal{G}^i|$. 
The server will aggregate the local models of chosen malicious clients and 
normally trained benign clients. 
The final backdoored graph classifier, denoted as $f_B$, is shared with all clients.  
During testing, malicious clients will use the same 
$\kappa$ for their testing graphs,  
but the trigger location is randomly chosen.  

\vspace{+0.2mm}
\noindent {\bf Rand-GDBA:}  
Each malicious client $i$ has its own local trigger $\kappa^i$ (often sparser/smaller than $\kappa$), and injects $\kappa^i$ into a fraction of its training graphs $\mathcal{G}^i$, where the trigger location is randomly chosen. Then each malicious client generates its backdoored graphs $\mathcal{G}^i_B = \{R({G}_{j}^{i},\kappa^i), y_B\}$ for training (i.e., minimizing the loss in Equation (\ref{equation:Rand-GCBA})). 
During testing, all malicious clients' triggers $\{\kappa^i\}$ will be combined into a single one. The combined trigger, with a random location, will be injected into testing graphs.

Rand-GDBA is shown to be 
more effective than Rand-GCBA~\cite{xu2022more}. 
Figure \ref{fig:Rand-GCBA-Rand-GDBA-Opt-GDBA} shows example 
shared trigger in Rand-GCBA, local triggers in Rand-GDBA across clients, and triggers in our attack.  

\subsection{Threat Model}
\label{sec:threatmodel}

We aim to understand the robustness of FedGL 
from both the attacker's and defender's perspective. 
As an attacker, we expect to design an effective and stealthy DBA to FedGL during training. 
As a defender, in contrast, we expect to design an effective certified defense against the worst-case DBA on a  backdoored FedGL model. 

%\vspace{+0.5mm} 
\noindent {\bf Attacker:} We assume the attacker manipulates a fraction (say $\rho$) of the total $C$ clients, namely malicious clients.   
\begin{itemize}[leftmargin=*]

% \vspace{+0.5mm} 
\item {\bf Attacker's knowledge:} All malicious clients only know their own training graphs and the shared global model in the whole of (backdoored) FedGL training. %each training iteration. 

%\vspace{+0.5mm} 
\item {\bf Attacker's capability:} 
Malicious clients can inject a subgraph trigger into any location/part of their training graphs during training. 
To ensure effectiveness and stealthiness for the attack, 
we follow~\cite{xie2019dba,xu2022more} to inject the trigger in every training iteration, but 
its size (w.r.t. number of nodes or/and edges) is small. 

%\vspace{+0.5mm} 
\item {\bf Attacker's objective:} 
Malicious clients aim to learn a backdoored FedGL model 
such that: it predicts the backdoored  testing graphs 
as the target label, while    
correctly predicting the clean testing graphs. 
This implies the model will achieve a \emph{high backdoor accuracy} as well as \emph{a high main task accuracy}.   
\end{itemize}

%\vspace{+0.5mm} 
\noindent {\bf Defender:} The defender aims to build a certifiably robust defense, under which the learnt backdoored FedGL 
can achieve two goals.  
\begin{itemize}[leftmargin=*]
\item {\bf High certified main task accuracy:} 
provably predict correct labels as many as possible 
for clean testing graphs against arbitrary trigger (any shape and location) with a bounded size.

\item {\bf Low certified backdoor accuracy:} provably predict the target label as few as possible 
for backdoored testing graphs (that are generated by our Opt-GDBA). 
\end{itemize}

\begin{figure*}[t]
\centering	
\includegraphics[width=0.95\textwidth]{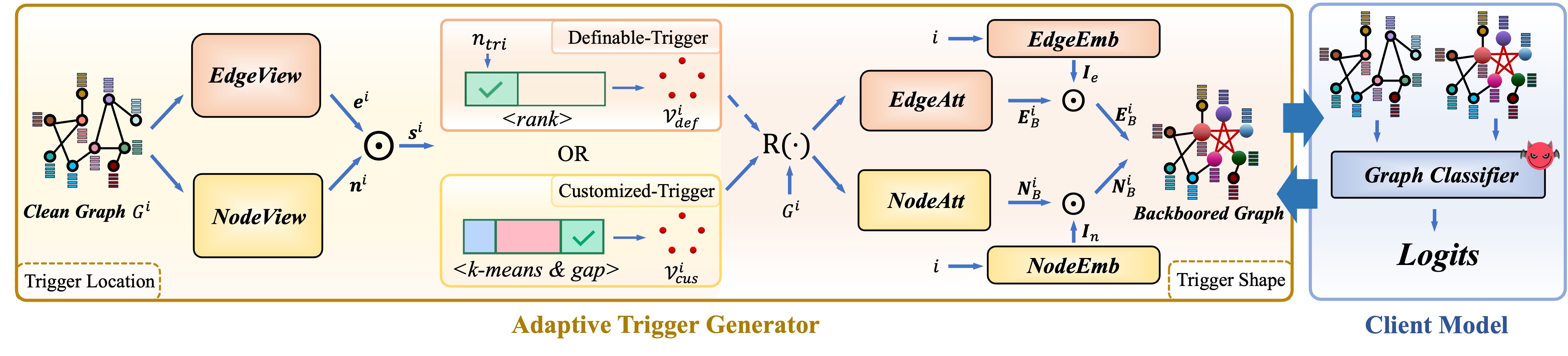} 
\vspace{-4mm}
\caption{Pipeline of our proposed Opt-GDBA on FedGL (a client $i$ perspective). 
}
\label{fig:Overview1}
\end{figure*}

%\vspace{-1mm}
\section{Optimized DBAs on FedGL}
\label{sec:attack}

Recall the existing DBA to FedGL generates triggers with \emph{random} locations and \emph{random} shape, and obtains unsatisfactory backdoor performance. 
We propose an optimized DBA on FedGL (called Opt-GDBA) to address the limitation. 
Our Opt-GDBA designs an adaptive trigger generator to {adaptively optimize the trigger location and shape} by integrating the edge and node feature information in individual graphs. See Figure \ref{fig:Overview1} for the pipeline of Opt-GDBA. 

\subsection{Adaptive Trigger Generator}
\label{sec:generator}

The proposed adaptive trigger generator consists of 
three modules: \emph{1) node importance score learning, 
2) trigger location learning, and 3) trigger shape learning}. For simplicity, we use a client $i$'s graph $G^i = (\mathcal{V}^i, \mathcal{E}^i, {\bf X}^i)$ with the adjacency matrix ${\bf A}^i$ 
for illustration. 
We first obtain the nodes' importance scores using module 1).  We next input the nodes' scores to module 2) to decide the trigger location with a predefined or customized trigger size. 
Finally, module 3) learns the trigger shape and generates the backdoored graph $G^i_B$ for $G^i$ together with module 2). 
{\bf \emph{Detailed architecture of all the described networks below are deferred in Appendix~\ref{app:netarch}.}}

\vspace{+0.5mm}
\noindent{\bf 1) Node importance score learning.} 
The goal is to measure the node importance  
so that the trigger can be placed on the important nodes. 
Specifically, we leverage both the edges and node features in $G^i$ to  decide the node importance. 

First, we define two networks: $\mathrm{EdgeView}(\cdot)$ and  $\mathrm{NodeView}(\cdot)$.  $\mathrm{EdgeView}(\cdot)$ 
characterizes the node importance from the edge view, and  
the extracted node importance scores 
from ${\bf A}^i$ 
are denoted by $ {\bf e}^{i} \in \mathbb{R}^{|\mathcal{V}^i|}$. 
Instead, $\mathrm{NodeView}(\cdot)$ characterizes the node importance from the node feature view,  
and the extracted node importance scores from ${\bf X}^i$ 
are denoted by ${\bf n}^{i} \in \mathbb{R}^{|\mathcal{V}^i|}$. 
Formally, 
\begin{equation}
{\bf e}^{i}  =  \mathrm{EdgeView}({\bf A}^{i} ), \ \ \ \  {\bf n}^{i}  =  \mathrm{NodeView}({\bf X}^{i} ),
\end{equation}
where ${\bf e}^{i}$ and ${\bf n}^{i}$ are constrained to have a value range $(0,1)$. 

We then calculate the nodes' importance scores, denoted by ${\bf s}^{i} $, as the element-wise product $\odot$ of vectors $ {\bf e}^i$ and $ {\bf n}^i$ as below: 
\begin{equation}
\begin{aligned} 
& {\bf s}^{i}  = {\bf e}^i \odot {\bf n}^i. 
\end{aligned} 
\label{equation: Trigger-location}
\end{equation}

%\vspace{+2mm}
\begin{algorithm}[h]%h
\caption{$k$-means\_gap to learn customized trigger size} 
\footnotesize
\label{algorithm: Customized-Trigger}
\begin{flushleft}
{\bf Input:} Nodes' scores ${\bf s}^{i}$, maximum trigger size $n^*_{tri}$. 

{\bf Output:} Important nodes $\mathcal{V}^i_{cus}$.
\end{flushleft}
\begin{algorithmic}[1]
\State ${{\bf s}^{i} } = {\bf s}^{i} /\left \| {\bf s}^{i}  \right \|_1$ 
\For{$k = 1,2,...,K$} 
    \State $C_i$, $\mu_i$ $=$ $k$-means (${\bf s}^{i} $) 
    \State $V_k =  {\textstyle \sum_{i=1}^{k}}  {\textstyle \sum_{x_j \in C_i}}\left \|x_j-\mu_j  \right \|^2  $
    \For{$b=1,2,...,B$}
        \State $x_{j,b}$ = sample($|{{\bf s}^{i} }|$) in $[0,1]$
        \State $C_{i,b}$, $\mu_{i,b}$ $=$ $k$-means ($x_{j,b}$)
        \State $V_{kb}^* =  {\textstyle \sum_{i=1}^{k}}  {\textstyle \sum_{x_{jb}\in C_{ib}}}\left \| x_{jb}-\mu_{ib} \right \|^2  $
    \EndFor
    \State $Gap(k)=\frac{1}{B} {\textstyle \sum_{i=1}^{B}}\log_{}{(V_{kb}^*)}-\log(V_k)$
    \State $v'=\frac{1}{B}  {\textstyle \sum_{i=1}^{B}}\log(V_{kb}^*) $
    \State $sd(k)=(\frac{1}{B} {\textstyle \sum_{i=1}^{B}}(\log(V_{kb}^*)-v')^2  )^{\frac{1}{2} }$
    \State $s_k'=\sqrt{\frac{1+B}{B}sd(k) } $
\EndFor
\State $\hat{k}=$ $min(k) \ s.t.\ Gap(k)-Gap(k+1)+s_{k+1}' \ge 0$
\State $C_i$, $\mu_i$$=\hat{k}$-means(${\bf s}^{i} $)
\State $\mathcal{V}^i_{cus}$ $=$ \{$C_i | $max\{Avg$(C_1), $Avg$(C_2), ..., $Avg$(C_{\hat{k}})$\}\}
\If{$|\mathcal{V}^i_{cus}| > msize$} 
\State $\mathcal{V}^i_{cus}$ $=$ sort($\mathcal{V}^i_{cus}$) 
\State $\mathcal{V}^i_{cus}$ = $\mathcal{V}^i_{cus}[:msize]$
\EndIf

\end{algorithmic}
\end{algorithm} 

\setlength{\textfloatsep}{2mm}

% \vspace{+0.5mm}
\noindent{\bf 2) Trigger location learning.} 
With nodes' importance scores, we design two schemes to decide the trigger location in each graph: \emph{Definable-Trigger} and \emph{Customized-Trigger}. 

\emph{Definable-Trigger:} It \emph{predefines} a trigger node size $n_{tri}$ used by all  backdoored graphs. 
Specifically, this scheme first ranks ${\bf s}^{i} $ in a descending order and selects the nodes  $\mathcal{V}^i_{def}$ from $G^{i}$ with the top $n_{tri}$ 
 values as the trigger location. 

\emph{Customized-Trigger:} One drawback of  
Definable-Trigger 
is that all backdoored graphs  use the same trigger size, but the graph size varies in practice. This would cause $\mathcal{V}^i_{def}$ misses important nodes if $G^{i}$ is a 
large graph but $n_{tri}$ is small, or includes non-important nodes if $G^{i}$ is a  
small graph but $n_{tri}$ is large.  
To address it, we further develop the Customized-Trigger scheme, which automatically \emph{learns} the best local trigger size of each graph during the FedGL training. The learnt most important nodes for $G^i$ are stored in $\mathcal{V}^i_{cus}$.

Our main idea is to adopt the Gap statistics~\cite{tibshirani2001estimating} based on $k$-means clustering\footnote{K-mean is a widely-adopted clustering algorithm that is efficient and effective. By integrating with gap statistics, K-means can also efficiently determine the optimal number of clusters. We admit there are other more advanced/complicated/effective clustering algorithms. Note that our purpose is not to pick the best clustering algorithm, but the one that is suitable to achieve our goal, i.e., learning the local trigger size.}. The algorithm details are shown in Algorithm \ref{algorithm: Customized-Trigger}. 
At a high-level, given the node scores ${\bf s}^{i} $, we first use the Gap statistics to estimate the number of clusters $\hat{k}$. Then we employ $k$-means to divide the nodes into $\hat{k}$ clusters based on their scores ${\bf s}^{i} $. Finally, the nodes  in the cluster with the largest average score are treated as the most important nodes $\mathcal{V}^i_{cus}$, whose positions are put the local trigger. 
\emph{Note that, to ensure the stealthiness of the attack, we require the trigger size not exceed a threshold (e.g., $n^*_{tri}$).} 

%\vspace{+0.5mm}
\noindent{\bf 3) Trigger shape learning.} Given the location of the local trigger ($\mathcal{V}^i_{def}$ or $\mathcal{V}^i_{cus}$), we learn the trigger shape through two sub-modules: \emph{edge/node attention}, and \emph{local trigger differentiation}. The former determines the edges and node features in the trigger,  which is inspired by \cite{xi2021graph}, while the latter promotes divergence of different local triggers so that the attack effectiveness can be enhanced when these local triggers are combined for backdoored testing. 
We denote $G_B^i = (\mathcal{V}^i, \mathcal{E}_B^i, {\bf X}_B^i)$ as an initial backdoored graph with 
an empty trigger shape and the corresponding adjacency matrix ${\bf A}^i_B$. 
For brevity, we use $\mathcal{V}^i_{def}$ for illustration. 

\emph{Edge/node attention.} 
We introduce two attention networks: \emph{edge attention network} $\mathrm{EdgeAtt}(\cdot)$ that focuses on understanding 
the connectivity between nodes $\mathcal{V}^i_{def}$ in the trigger, and \emph{node attention network} $\mathrm{NodeAtt}(\cdot)$ that aims to improve the flexibility of the trigger by also incorporating node features.  
We denote the trigger's edge attention matrix as ${\bf E}_{tri}^i  \in \mathbb{R}^{|\mathcal{V}^i_{def}| \times |\mathcal{V}^i_{def}|}$, and trigger's node feature attention matrix as ${\bf N}_{tri}^i  \in \mathbb{R}^{|\mathcal{V}^i_{def}| \times d}$. 
Formally, 
\begin{align}
    {\bf E}_{tri}^i =\mathrm{EdgeAtt}({\bf A}^i_B, \mathcal{V}^i_\text{def}); \label{equation: Trigger-shape-edge} \\
     {\bf N}_{tri}^i =\mathrm{NodeAtt}({\bf X}^i_B, \mathcal{V}^i_\text{def}). \label{equation: Trigger-shape-node} % 
\end{align} 

\begin{algorithm}[!t]%h
\small
\caption{Adaptive Trigger Generator} 
\label{algorithm: Generator}
\begin{flushleft}
{\bf Input:} A clean graph $G^{i}$, trigger node size $n_{tri}$ or $n^*_{tri}$. 

{\bf Output:} Backdoored graph $\tilde{G}^{i}_B $.  
\end{flushleft}
\begin{algorithmic}[1]
\State  ${\bf s}^{i} = $ $\mathrm{Node\_score}$($G^{i} $) // Node importance score
\If{Definable-Trigger}
    \State $\mathcal{V}^i_{def}$ $=$ rank(${\bf s}^{i} $, $n_{tri}$) // Trigger location 
    \State  $\tilde{G}^{i}_B = \mathrm{Trigger\_shape\_learning}$($G^{i} $, $\mathcal{V}^i_{def}$) 
\ElsIf{Customized-Trigger}
    \State $\mathcal{V}^i_{cus}$ $=$ $k$-means\_gap(${\bf s}^{i}, msize$) //  Trigger location
    \State  $\tilde{G}^{i}_B =$ $\mathrm{Trigger\_shape\_learning}$($G^{i} $, $\mathcal{V}^i_{cus}$)
\EndIf
\end{algorithmic}
\end{algorithm}

\setlength{\textfloatsep}{2mm}

\emph{Local trigger differentiation.}
To further enable distinct malicious clients to possess personalized and controllable local triggers, we also propose to incorporate the client index $i$ into the trigger shape generation. Specifically, we first use an edge embedding function $\mathrm{EdgeEmb}(\cdot)$ to convert the client index $i$ into ${\bf I}_e \in \mathbb{R}^{|\mathcal{V}^i_{def}| \times |\mathcal{V}^i_{def}|}$ and a node embedding function $\mathrm{NodeEmb}(\cdot)$ to convert it into ${\bf I}_n \in \mathbb{R}^{|\mathcal{V}^i_{def}| \times d}$. We then multiply ${\bf I}_e$ (and ${\bf I}_n$) with the attention matrix ${\bf E}_{tri}^i $ (and ${\bf N}_{tri}^i $) to integrate the unique information of the client index. 
The equations can be expressed as follows:
\begin{equation}
\begin{aligned}
&{\bf I}_e = \mathrm{EdgeEmb}(i), \ \ \ \ {\bf E}_{tri}^i ={\bf E}_{tri}^i \odot {\bf I}_e,  \\ 
& {\bf I}_n = \mathrm{NodeEmb}(i), \ \ \ \ {\bf N}_{tri}^i={\bf N}_{tri}^i \odot {\bf I}_n.
\end{aligned} 
\label{equation: Trigger-shape-different-local-triggers}
\end{equation}
To further discretize the connectivity status between nodes, we convert the continuous edge attention matrix ${\bf E}_{tri}^i$ to be binary, i.e., ${\bf E}_{tri}^i = \mathbbm{1}({{\bf E}_{tri}^i \ge 0.5}$), where $\mathbbm{1}(p)$ is an indicator function that returns $1$ if $p$ is true, and 0 otherwise. 

The trigger location $\mathcal{V}^i_{\textrm{def}}$ as well as trigger shape ${\bf E}_{tri}^i, {\bf N}_{tri}^i$ 
decides the optimized trigger, which we denote as $\tilde{\kappa}^i = (\mathcal{V}^i_{\textrm{def}}, {\bf E}_{tri}^i, {\bf N}_{tri}^i)$.  
The optimized backdoored graph for a graph $G^i$ is then generated by $\tilde{G}^{i}_B = R(G^{i}, \tilde{\kappa}^i)$. 
Algorithm \ref{algorithm: Generator} summarizes the  adaptive trigger generator for generating a backdoored graph. 

\begin{algorithm}[!t]%h
\caption{Backdoored FedGL training with Opt-GDBA} 
\label{algorithm: OptGDBA}
\small
\begin{flushleft}
{\bf Input:} Total clients $\mathcal{C}$ with clean graphs $\{\mathcal{G}^i\}_{i\in\mathcal{C}}$, malicious clients $\tilde{\mathcal{C}}$, training iterations $iter$,  initial global model $\theta_1$,  malicious clients' initial generator model $\{\omega^i_1\}_{i \in \tilde{\mathcal{C}}}$. 

{\bf Output:} Backdoored global model ${\theta}_{iter}$.
\end{flushleft}
\begin{algorithmic}[1]
\For{each iteration $t$ in [1,iter]}
    \For{each client $i\in \mathcal{C}$} 
        \If{$i \in \tilde{\mathcal{C}}$}:
            \State Client $i$ divides $\mathcal{G}^i$ into  $\mathcal{G}^i_C$ and to-be-backdoored  $\mathcal{G}^i_o$ 
            \State $\tilde{\mathcal{G}}^i_B = \mathrm{Generator}(\mathcal{G}^i_{o};\omega_t^{i})$ using Algorithm~\ref{algorithm: Generator}
            \State $\theta^{i}_t =  \underset{\theta^{i}}{\arg\min} \,  L(\tilde{\mathcal{G}}^i_B \cup \mathcal{G}^i_C;{\theta}_t)$

            \State $\omega^{i}_t =  \underset{\omega^{i}}{\arg\min} \,  L(\tilde{\mathcal{G}}^i_B;\theta^{i}_t  )$
        \Else
            \State $\theta^{i}_t = \underset{\theta^{i}}{\arg\min} \, L(\mathcal{G}^i;{\theta}_t)$ 
        \EndIf
    \EndFor
    
    \State Server randomly selects $\mathcal{C}_t$ clients for aggregation
    \State $\theta_{t+1} =\frac{1}{|\mathcal{C}_t|}  {\textstyle \sum_{i \in \mathcal{C}_t}} \theta^{i}_t$
\EndFor
\end{algorithmic}
\end{algorithm}

\subsection{FedGL Training with Optimized Backdoor}
\label{sec:discriminator}

We now show the entire backdoored FedGL training with the optimized backdoored graphs (algorithm details are in Algorithm~\ref{algorithm: OptGDBA}). It involves alternatively and iteratively training the (backdoored) local  model, optimizing the adaptive trigger generator, and updating the shared global backdoored model.

\vspace{+0.5mm}
\noindent {\bf Training the (backdoored) local model:} 
We denote the optimized backdoored graphs in each malicious client $i$ as $\tilde{\mathcal{G}}^i_B = \{R({G}_{j}^{i},\tilde{\kappa}_{j}^i), y_B\}$.   
Then each malicious client $i$ trains its local backdoored model ${\theta}^{i}_B$ via minimizing the loss on both the optimized backdoored graphs $\tilde{\mathcal{G}}^i_B$ and clean graphs $\mathcal{G}^i_C$: 
\begin{align} 
 & {\theta}^{i}_B =  
 {\arg\min}_{{\theta}^{i}_B} \,  L(\tilde{\mathcal{G}}^i_B \cup \mathcal{G}^i_C; \theta).   
 \label{eqn:opt-malmodel}
\end{align}
For each benign client $j$, it updates the local model via minimizing the loss on all its clean graphs $\mathcal{G}^j$ as:
\begin{align} 
 {\theta}^{j} =  
 {\arg\min}_{\theta^{j}} \,  L(\tilde{\mathcal{G}}^j; \theta).  
 \label{eqn:opt-benmodel}
\end{align}

\vspace{+0.5mm}
\noindent {\bf Optimizing the adaptive trigger generator:} 
We denote the parameters of the trigger generator per malicious client $i$ as $\omega^i$, which includes  the  parameters of all networks in Section~\ref{sec:generator}. 
Each malicious client $i$ also 
optimizes its generator $\omega^i$ to ensure the generated backdoored graphs be more effective and diverse. Specifically, 
\begin{align} 
\omega^i =  
{\arg\min}_{\omega^i} \, L(\tilde{\mathcal{G}}^i_B; {\theta}^i_B).  
 \label{eqn:opt-generator}
\end{align}

\vspace{+0.5mm}
\noindent {\bf Updating the shared global model:} The server averages the backdoored local models $\{{\theta}^{i}_B\}$ and benign local models $\{{\theta}^{j}\}$ of the selected clients to update the global model $\theta$.

\begin{table*}[!t]
%\footnotesize
\small
\addtolength{\tabcolsep}{1pt}
\caption{Results of all the compared attacks in the default setting. The gain is between Opt-GDBA and Rand-GDBA. 
}
% \vspace{-4mm}
\centering
\begin{tabular}{c||c c c|c|c c|c c|c c}
\toprule
{\bf Datasets} & \multicolumn{3}{c}{{\bf Customized-Trigger} ($n^*_{tri}=5$)} &  \multicolumn{3}{c}{{\bf Opt-GDBA}} & \multicolumn{2}{c}{{\bf Rand-GDBA}} & \multicolumn{2}{c}{\bf Rand-GCBA} \\
%\cline{2-11}
{\bf (MA without attack)} & (MA\ /\ BA) & $n_{tri}$ & $e_{tri}$ & $n_{tri}$ & (MA\ /\ BA) & $e_{tri}$ & (MA\ /\ BA) & $e_{tri}$ & (MA\ /\ BA) & $e_{tri}$ \\ % 
\hline
\hline
BITCOIN (MA=0.73) & 0.72\ /\ 0.99 (\textcolor{red}{$\uparrow$0.36})& 4.29 & 4.20 & 4 & 0.72\ /\ 0.99 (\textcolor{red}{$\uparrow$0.36}) & 3.08 & 0.71\ /\ 0.63 & 4 & 0.72\ /\ 0.57 & 6\\
%\hline
MUTAG  (MA=0.74) & 0.72\ /\ 0.95 (\textcolor{red}{$\uparrow$0.43})& 3.51 & 2.31 & 4 & 0.71\ /\ 0.85 (\textcolor{red}{$\uparrow$0.33}) & 3.41 & 0.71\ /\ 0.52 & 4 & 0.73\ /\ 0.48 & 6\\
%\hline
PROTEINS  (MA=0.73) & 0.72\ /\ 0.90 (\textcolor{red}{$\uparrow$0.39}) & 3.75  &  2.56 & 4 & 0.72\ /\ 0.90 (\textcolor{red}{$\uparrow$0.39}) & 2.07  &  0.70\ /\ 0.51 & 4 & 0.71\ /\ 0.33  & 6\\
%\hline
DD  (MA=0.73) & 0.72\ /\ 0.86 (\textcolor{red}{$\uparrow$0.46}) &  3.19 & 1.57  & 4 &  0.72\ /\ 0.78 (\textcolor{red}{$\uparrow$0.38}) & 2.97  & 0.72\ /\ 0.40  & 4 &  0.72\ /\ 0.33 & 6\\
%\hline
COLLAB  (MA=0.75) & 0.73\ /\ 0.86 (\textcolor{red}{$\uparrow$0.32}) & 4.68 & 4.51 & 4 & 0.73\ /\ 0.84 (\textcolor{red}{$\uparrow$0.30}) & 3.34 & 0.73\ /\ 0.54 & 4 & 0.71\ /\ 0.37 & 6\\
%\hline
RDT-M5K  (MA=0.53) & 0.52\ /\ 0.90 (\textcolor{red}{$\uparrow$0.33}) & 4.51  & 3.59  & 4 & 0.52\ /\ 0.89 (\textcolor{red}{$\uparrow$0.32}) & 3.27  & 0.52\ /\ 0.57 & 4 & 0.52\ /\ 0.40  & 6\\
\bottomrule
%---------------------------------------------------------
\end{tabular}
\label{table:GraphDBA-results}  
\vspace{-2mm}
\end{table*}

\section{Attack Results}
\label{sec:attackresults}

\subsection{Experimental Setup}
\label{sec:setup}

%\vspace{-2mm}
\noindent{\bf Datasets and training/testing sets:} We evaluate our attack on six benchmark 
real-world graph datasets for graph classification. 
{\bf \emph{Dataset description and statistics and training/testing sets about the datasets are presented in Table \ref{table:datasets} in Appendix~\ref{app:dataset}.}} 

\vspace{+0.5mm}
\noindent{\bf Attack baselines:} 
We compare our Opt-GDBA with Rand-GCBA and Rand-GDBA (details are in Section~\ref{sec:BAFedGL}). Their main difference lies in the way to inject the trigger.

\begin{itemize}[leftmargin=*]
\item {\bf Rand-GCBA~\cite{xu2022more}:} All malicious clients use a shared trigger with the same shape but random location. 
To force the trigger yields the most effective attack, we assume it be a \emph{complete subgraph}.  

\item {\bf Rand-GDBA~\cite{xu2022more}:} Each malicious client generates its local
trigger. 
Following~\cite{zhang2021backdoor}, each client generates the trigger using the Erdős-Rényi (ER) random graph model~\cite{gilbert1959random}, where the number of edges $e_{tri}$ with a trigger node size $n_{tri}$ can be controlled.  
These triggers are then attached to random nodes in the to-be-backdoored  
graphs. 
{To further enhance this attack, we also maintain the diversity of local triggers among malicious clients. To do so, we store a set of generated local triggers via the ER model, and assign different triggers to different malicious clients.}

For fair comparison, we make sure the total number of edges in all triggers in Rand-GCBA and Rand-GDBA are same. This can be realized by forcing $\rho^{c} * e_{tri}^{c} = \rho^{d} * e_{tri}^{d}$, where $\rho^c$ and $\rho^d$ are the ratio of malicious clients, and $e_{tri}^{c} $ and $e_{tri}^{d}$ are the number of trigger edge in Rand-GCBA and Rand-GDBA, respectively.

\item {\bf Our Opt-GDBA:} Each to-be-backdoored training graph generates an individual trigger via the proposed adaptive trigger generator, where the trigger location and shape are learnt. Note that the trigger node size $n_{tri}$ is predefined in the Definable-Trigger scheme (same as Rand-GCBA and Rand-GDBA), but is automatically learnt in the Customized-Trigger scheme.  
% \vspace{-2mm}
\end{itemize}

During testing, the local triggers are combined into a global trigger. 
For fair comparison, we let all attacks use a {complete subgraph} as the global trigger.
In our  Opt-GDBA, 
for each testing graph, we learn the important nodes that determine the trigger location, and then generate the complete graph based on them.  
In contrast, the trigger location of Rand-GCBA and Rand-GDBA in testing graphs is random. 
\emph{Note that \cite{xu2022more} uses a different way to combine local triggers and injects a much larger global trigger in Rand-GDBA.} {\bf \emph{The discussion and results are shown in Table~\ref{table:comparsion experiment} in Appendix~\ref{app:attackresults}}}.

\vspace{+0.5mm}
\noindent{\bf Parameter setting:} 
During FedGL training, we use a total of $C=40$ 
clients {($C=20$ on MUTAG due to less data)} and evenly distribute the training graphs in each dataset to the clients. The total number of iterations is  200 in all datasets, except the larger RDT-M5K that is 400. 
In each iteration, the server randomly selects $50\%$ of the total clients for training. The clients use the \emph{de facto} Graph Isomorphism Network (GIN)~\cite{xu2018powerful} as the graph classifier. 
We use its open-source code ({https://github.com/weihua916/powerful-gnns}) in our experiments. 
By default, 50\% of malicious clients' 
training graphs are randomly sampled to inject the backdoor trigger, 
and the target label is 1. 
All testing graphs are chosen for backdoored testing.  \\  
There are several hyperparameters that can affect all attacks' performance on FedGL: fraction of malicious clients $\rho$ and trigger node size $n_{tri}$ or a threshold size $n^*_{tri}$ in Customized-Trigger scheme. 
We set $\rho=20\%$ and $n_{tri}=4$ by default, and set  $n^*_{tri}=5$ in all experiments. 
We will also study the impact of these hyperparameters. 
Our Opt-GDBA contains many modules and we will also study the importance of individual module.

\vspace{+0.5mm}
\noindent{\bf Evaluation metrics:} 
We adopt four metrics for evaluation: the main task accuracy (MA) and backdoor accuracy (BA) on testing graphs; the average trigger node size (also use $n_{tri}$) and edge size (also use $e_{tri}$) of all backdoored training graphs. 
A more effective attack would achieve a higher MA and higher BA, and  
a more stealthy attack would have a lower $n_{tri}$ and $e_{tri}$ given a close MA/BA.

\begin{figure*}[!t]
\centering	
\includegraphics[width=0.95\textwidth]{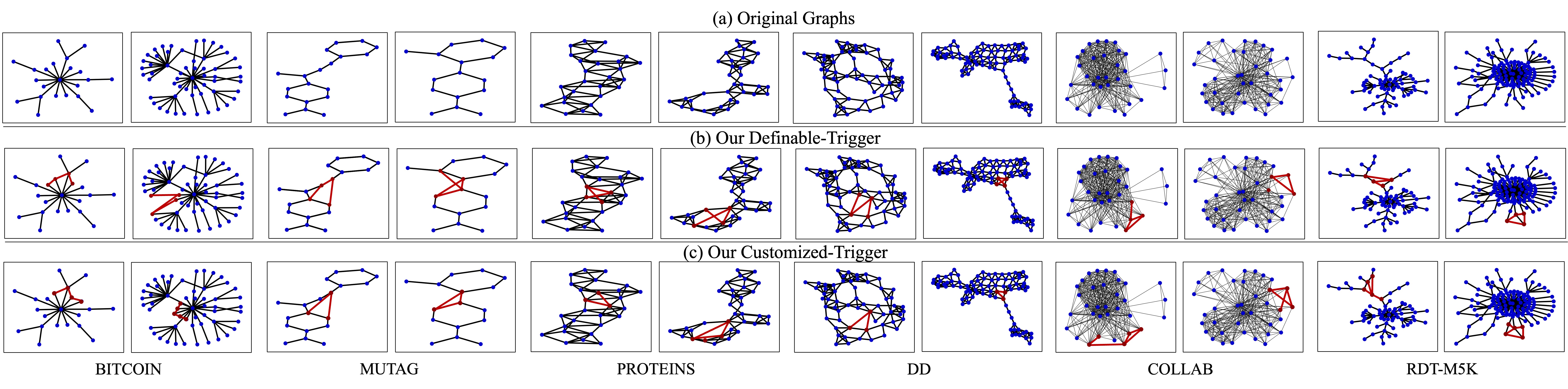}
\vspace{-2mm}
\caption{Examples of original clean graphs on the six datasets and their corresponding backdoored ones by our Opt-GDBA. 
}
\label{fig:triggersamples}
% \vspace{-4mm}
\end{figure*}

\begin{figure*}[!t]
\centering	
\includegraphics[width=\textwidth]{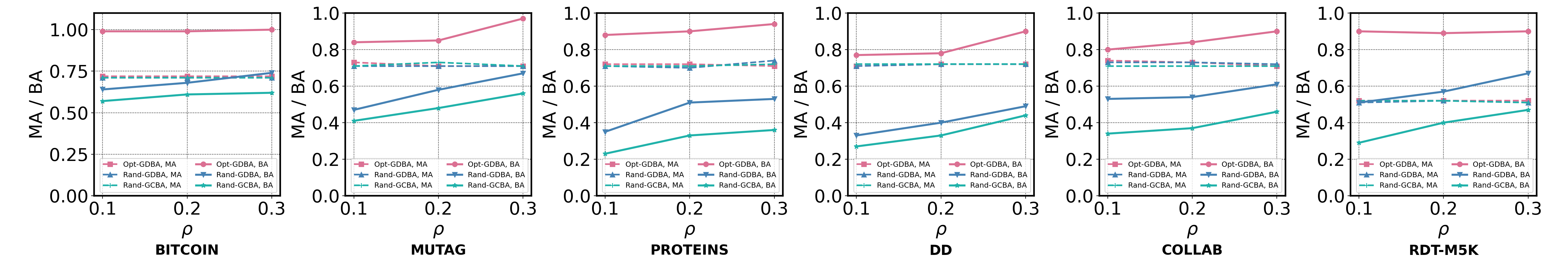}
% \vspace{-8mm}
\caption{MA/BA vs. $\rho$ on all compared attacks in all datasets.} 
%\vspace{-4mm}
\label{fig:attack-f}
% \vspace{-3mm}
\end{figure*}

\begin{figure*}[!t]
\centering	
\includegraphics[width=\textwidth]{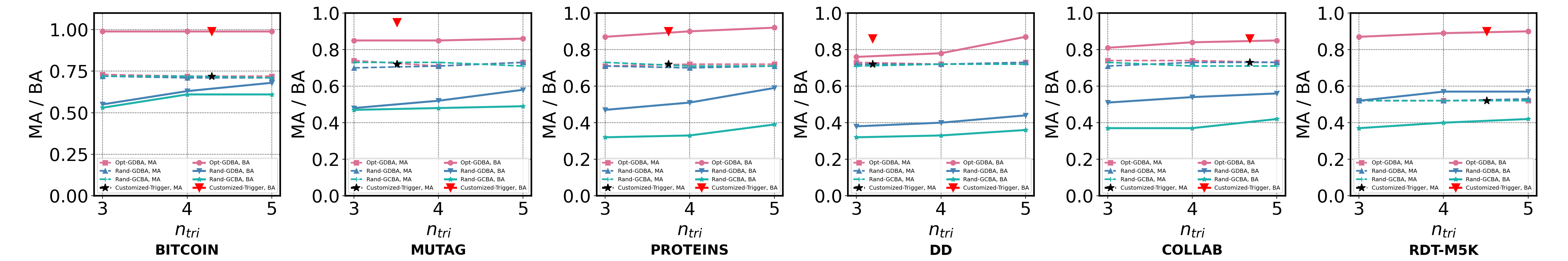}
% \vspace{-8mm}
\caption{MA/BA vs. $n_{tri}$ ($n^*_{tri}=5$) on all compared attacks in all datasets.} 
\label{fig:attack-size}
% \vspace{-3mm}
\end{figure*}

\begin{table*}[!t]
\footnotesize
\renewcommand{\arraystretch}{1}
%\addtolength{\tabcolsep}{-2pt}
\caption{Comparing the two trigger location learning schemes in our Opt-GDBA 
w.r.t. the average trigger edge size $e_{tri}$.  
}
% \vspace{-4mm}
\centering
    \begin{tabular}{c||c c c|c c c|c c c|c c c|c c c|c c c }
    \toprule
    {\bf Datasets} & \multicolumn{3}{c}{BITCOIN} & \multicolumn{3}{c}{MUTAG} & \multicolumn{3}{c}{PROTEINS} & \multicolumn{3}{c}{DD} & \multicolumn{3}{c}{COLLAB} & \multicolumn{3}{c}{RDT-M5K} \\
    $\rho$ vs. $e_{tri}$ & 10\% & 20\% & 30\% & 10\% & 20\% & 30\% & 10\% & 20\% & 30\% & 10\% & 20\% & 30\% & 10\% & 20\% & 30\% & 10\% & 20\% & 30\%  \\
    % \hline
    \hline
    \hline
    {\bf Definable-Trigger} &  4.95 &   5.09 &  5.35 &  5.83 &   5.88 &   5.63 &   3.56 &   3.36 &   3.44 &   5.42 &   5.04 &   5.22 &   5.85 &   5.72 &   5.99 &   5.39 &   5.45 &     5.59 \\
    {\bf Customized-Trigger} &   3.88 &   4.20 &   4.35 &   3.52 &   2.31 &   2.32 &   2.68 &   2.56 &   2.54 &   1.64 &   1.57 &   1.79&   4.50 &   4.51 &   4.72 &   3.42 &   3.59 &   3.86  \\
    \bottomrule
    \end{tabular}
\label{table:cus-def}  
% \vspace{-2mm}
\end{table*}

%\vspace{-4mm}
\subsection{Experimental Results}
\label{5.2}
%\vspace{-2mm}

\subsubsection{Main results of the compared attacks} 
Table \ref{table:GraphDBA-results} shows the comparison results of the attacks in the default setting ({\bf \emph{more comprehensive comparison and more results are shown in Table \ref{table:GraphDBA-results-full}-Table~\ref{table: empirical defenses}  
in Appendix~\ref{app:attackresults})}}.  
We have the below key observations: 

\begin{enumerate}[leftmargin=*]

\vspace{+0.5mm}
\item {\bf Main task performance is marginally sacrificed under all attacks:} All attacks achieve a close MA, compared to the MA without attack (i.e., the differences between them in all cases are $\leq 3\%$). This verifies these attacks only slightly affect the performance of the main task.  

\vspace{+0.5mm}
\item {\bf Rand-GDBA outperforms Rand-GCBA on attacking FedGL:} Similar to the conclusion of backdoor attacks on image data~\cite{xie2019dba},  distributed backdoors for graph data are also superior to centralized backdoors on attacking FedGL. Specifically, the BA of Rand-GDBA is higher than that of Rand-GCBA, with a gain from $4\%$ to $18\%$. 
This implies the diverse local triggers indeed can promote the backdoor attack.  
This observation hence implies the importance of using distributed backdoors. 

\vspace{+0.5mm}
\item {\bf Our Opt-GDBA outperforms Rand-GDBA in terms of both attack effectiveness and stealthiness:} 
Both our Opt-GDBA with Definable-Trigger  and  Customized-Trigger schemes
 yield impressive
attack results, with BA exceeding 85\% and 90\% in almost all cases, respectively. 
Under a same trigger node size $n_{tri}$ (Definable-Trigger) or a smaller learnt $n_{tri}$ (Customized-Trigger), the BA of Opt-GDBA consistently and significantly surpasses that of Rand-GDBA. Particularly, the gain is  
from $30\%$ to $46\%$. 
In addition, the average number of injected edges of Opt-GDBA 
is less than that of Rand-GDBA, showing Opt-GDBA is a more stealthy attack than Rand-GDBA. 
These findings underscore that the graph-dependent triggers 
learnt by our trigger generator are not only better memorized during the FedGL training, but also  uncover important locations in the clean graphs. 
Figure~\ref{fig:triggersamples} shows example triggers generated by Opt-GDBA on the six datasets. We can see most of the triggers are attached to the important/central nodes in the raw graphs.  
\vspace{-2mm}

\end{enumerate}

\begin{figure*}[!t]
\centering	
\includegraphics[width=0.85\textwidth]{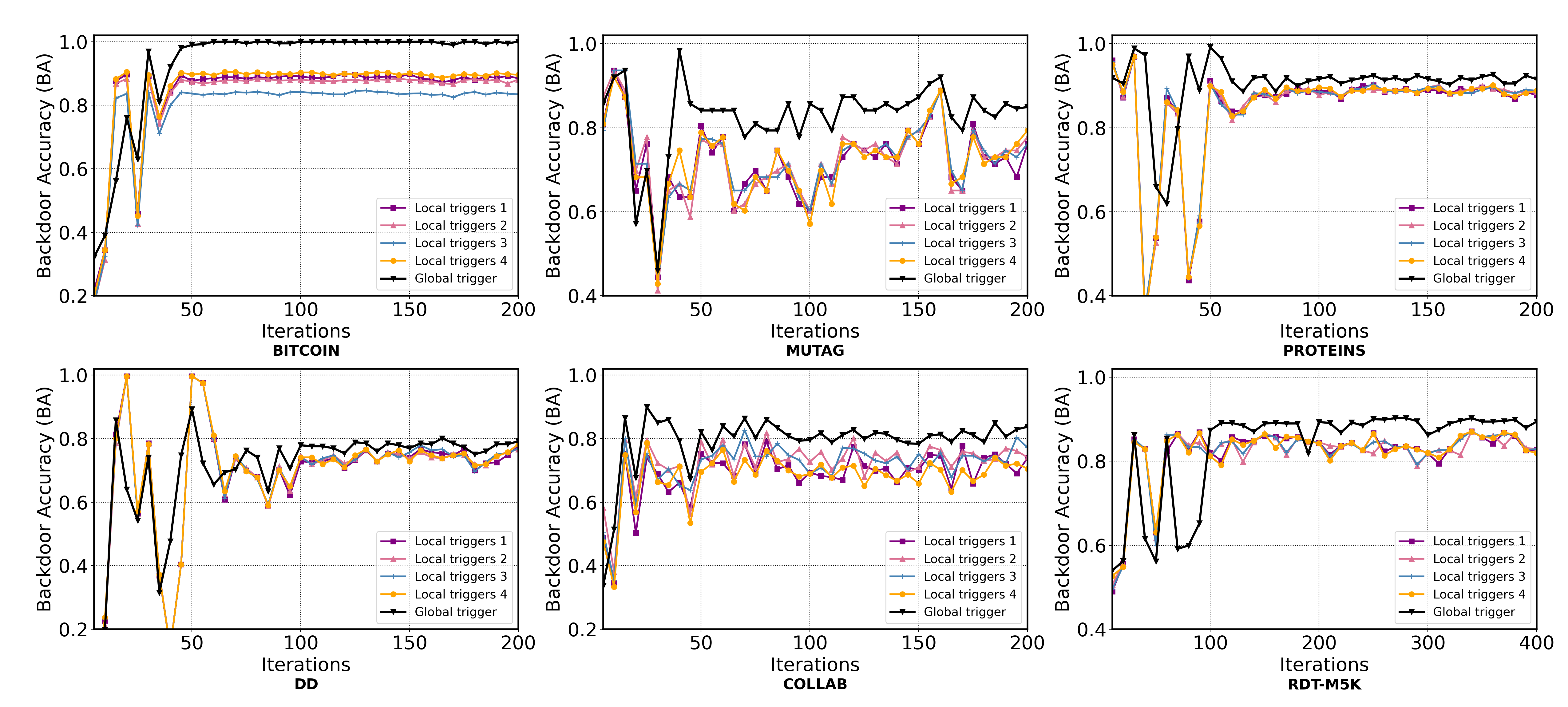}
\vspace{-4mm}
\caption{Comparing the backdoor performance with global trigger vs. local triggers generated by malicious clients.  
}
\label{fig:L-trigger-vs-G-trigger}
%\vspace{-2mm}
\end{figure*}

\begin{table*}[!t]
\footnotesize
\renewcommand{\arraystretch}{1.0}
\addtolength{\tabcolsep}{-1.5pt}
\caption{Structure similarity between the generated backdoored graphs by our Opt-GDBA and the clean graphs.  
}
% \vspace{-4mm}
\centering
    \begin{tabular}{c||c c c c|c c c c|c c c c|c c c c|c c c c|c c c c}
    \toprule
    {\bf Datasets} & \multicolumn{4}{c}{BITCOIN} & \multicolumn{4}{c}{MUTAG} & \multicolumn{4}{c}{PROTEINS} & \multicolumn{4}{c}{DD} & \multicolumn{4}{c}{COLLAB} & \multicolumn{4}{c}{RDT-M5K} \\
    % \hline
    $n_{tri} (n^*_{tri})$& 3 & 4 & 5 & $5^*$ & 3 & 4 & 5 & $5^*$ & 3 & 4 & 5 & $5^*$ & 3 & 4 & 5 & $5^*$ &  3 & 4 & 5 & $5^*$ & 3 & 4 & 5 & $5^*$ \\
    \hline
    \hline
    NetSim ($\uparrow$) & 0.73  &  0.55 & 0.52 & 0.54 & 0.99 &   0.90 & 0.82 & 0.87 & 0.93  &  0.88 & 0.80 & 0.86 &1.00 &  0.99 & 0.99 & 0.99 &1.00 &  0.99 & 0.99 & 0.99 &0.99 &  0.98 & 0.97&  0.98 \\
    DeltaCon ($\uparrow$) & 0.80 &  0.65 & 0.63 & 0.64 &0.96  &   0.93 & 0.89 & 0.92 &  0.95 &   0.94  & 0.89 & 0.91 & 1.00 &  0.99 & 0.99 & 1.00 &1.00 &  0.99 & 0.99 & 0.99 & 0.99 &  0.99 & 0.98& 0.98\\
    \bottomrule
    \end{tabular}
\label{table:metric}  
 \vspace{-2mm}
\end{table*}

% \vspace{-2mm}
\subsubsection{Impact of hyperparameters on our Opt-GDBA} 
In this set of experiments, we will study in-depth the impact of the important hyperparameters 
on our Opt-GDBA. 

\vspace{+0.5mm}
\noindent {\bf Impact of the fraction $\rho$ of malicious clients:}
Figure~\ref{fig:attack-f} shows the MA/BA results vs. 
$\rho=10\%,20\%,30\%$. 
We can see MA is stable w.r.t.  
different $\rho$, and 
BA (slightly) increases with a larger $\rho$. For instance, on MUTAG, when 
$\rho$ is from $10\%$ to $30\%$, the BA of Rand-GCBA, Rand-GDBA, and 
our Opt-GDBA can be increased from 41\% to 56\%, from 43\% to 67\%  and from $84\%$ to $97\%$  with the Definable-Trigger, respectively.   
This shows MA is marginally affected, but the attack becomes stronger with more malicious clients.   

\vspace{+0.5mm}
\noindent {\bf Impact of the trigger size $n_{tri}$:} 
Figure~\ref{fig:attack-size} shows the MA/BA results vs. $n_{tri}$ (=3,4,5).  
We can see a larger trigger size corresponds to a larger BA, which implies a stronger attack. This is because the trigger can be injected to a larger region of the clean graph. Still, the MA 
is the very stable in terms of different trigger sizes.

\vspace{+0.5mm}
\noindent {\bf Impact of the trigger location learning scheme:}  
Our Opt-GDBA uses two schemes to decide the trigger location: Customized-Trigger  automatically learns it, while Definable-Trigger predefines it. 
Table~\ref{table:cus-def} shows the comparison results. 
We can see the trigger outputted by Customized-Trigger has an average number of 
edges $<5$ in all cases. 
In contrast,  Definable-Trigger yields $\gtrapprox 5$ edges  
in most cases. 
Note that the MA and BA of the two schemes are close. This hence reflects the Customized-Trigger scheme can further locate the ``more important'' region in a graph to attach the trigger. 

\vspace{+0.5mm}
\noindent{\bf Global trigger vs. local triggers:}
From Table \ref{table:GraphDBA-results}, we know the MA performance is marginally affected by the backdoor attacks with respect to different trigger sizes. 
Recall that, during testing, we use the combined local triggers to form a global trigger (a complete subgraph), which is injected into all testing graphs. In this experiment, we  
also explore the BA performance of our Opt-GDBA where each client uses the local triggers generated by its own trigger generator. 
Specifically, we use $\rho=20\%$ and the total number of malicious clients is $20\% * 50\%* 40=4$. 
Figure \ref{fig:L-trigger-vs-G-trigger} compares the BA  produced by the global trigger vs. local triggers per malicious client with Definable-Trigger Opt-GDBA with $n_{tri}=4$. 
For instance, ``Local triggers 1'' means the local triggers are generated via malicious client 1's trigger generator on the corresponding testing graphs.

We observe that though the backdoored FedGL training does not involve the global trigger, the BA achieved by the global trigger is even larger than that by the local triggers. One possible reason could be 
that the federated training might  memorize the combined effect of local triggers.  
This phenomenon further reinforces the FedGL framework is more  vulnerable to distributed backdoors.

\begin{table}[!t]
\footnotesize
\renewcommand{\arraystretch}{1.0}
\addtolength{\tabcolsep}{-1pt}
\caption{Finetuning the backdoored FedGL model by extending the training on clean graphs. 
}
% \vspace{-4mm}
\centering
    \begin{tabular}{c||c|c c c|c c c|c c c}
    \toprule
    {\bf Datasets} & \multicolumn{1}{c}{$\rho $} & \multicolumn{3}{c}{$10\%$} &  \multicolumn{3}{c}{$20\%$} &  \multicolumn{3}{c}{$30\%$}\\ 
    %\cline{2-11}
    & $n_{tri}$ & 3 &   4 & 5 &3 &   4 & 5 &3 &   4 & 5 \\
    \hline
    \hline
     \rowcolor{gray!15} \cellcolor{white}BITCOIN& BA & 0.97 & 0.99 & 0.98 & 0.99 & 0.99 & 0.99 & 0.99 & 1.00 & 0.95 \\
    %\cline{2-11}
    & BA-FT & 0.95 & 0.97 & 0.96 & 0.94 & 0.95 & 0.96 & 0.96 & 0.95 & 0.73\\
    %\hline
     \rowcolor{gray!15} \cellcolor{white} MUTAG & BA & 0.83 & 0.84 & 0.82 & 0.87 & 0.85 & 0.86 & 0.95 & 0.97 &0.94 \\
    %\cline{2-11}
    & BA-FT & 0.82 & 0.82 & 0.81 & 0.85 & 0.84 & 0.84 & 0.92 & 0.92 & 0.65\\
    %\hline
     \rowcolor{gray!15} \cellcolor{white}PROTEINS & BA & 0.82  & 0.91  &  0.88 & 0.87& 0.92  &0.90   & 0.94  & 0.94  & 0.96  \\
    %\cline{2-11}
    & BA-FT &  0.77 & 0.90  & 0.82  & 0.87 & 0.89  &  0.76 & 0.94  & 0.92  & 0.65  \\
    %\hline
     \rowcolor{gray!15} \cellcolor{white}DD& BA &  0.78 & 0.77  &  0.80 & 0.76 & 0.78  & 0.87  & 0.83  & 0.80  & 0.92  \\
    %\cline{2-11}
    & BA-FT & 0.70  & 0.74  &  0.76 & 0.72 & 0.76  &  0.84 & 0.75  & 0.73  &  0.61 \\
    %\hline
     \rowcolor{gray!15} \cellcolor{white}COLLAB& BA & 0.80  & 0.80  & 0.82  & 0.81 & 0.84  &0.85   & 0.85  & 0.90  & 0.91  \\
    %\cline{2-11}
    & BA-FT &  0.79 &  0.77 & 0.80  & 0.79 & 0.81  & 0.80  & 0.82  &  0.85 & 0.77  \\
    % \hline
     \rowcolor{gray!15} \cellcolor{white}RDT-M5K& BA & 0.88  & 0.90  & 0.85  & 0.87 & 0.89  & 0.90  & 0.90  & 0.89  & 0.92  \\
    %\cline{2-11}
    & BA-FT & 0.87  &  0.85 & 0.72 & 0.85 & 0.94  & 0.75  & 0.88  &  0.86 & 0.60  \\
    \bottomrule
    \end{tabular}
\label{table:BA-FT-by-graphs} 
%\vspace{-2mm}
\end{table}

\vspace{-2mm}
\subsubsection{Persistence and stealthiness of the triggers}
\label{5.5} 
In this experiment, we explore the persistence and stealthiness of the  backdoor triggers generated by our Opt-GDBA. 

\vspace{+0.5mm}
\noindent {\bf (Persistence) Finetuning the backdoored FedGL model with clean graphs cannot remove the backdoor effect:}  
A straightforward strategy to mitigate the backdoor effect is to finetune the FedGL model only with clean graphs. To simulate this, we extend the FedGL training with an extra 200 iterations (e.g., from 200 to 400) which only involves training with clean graphs. Table~\ref{table:BA-FT-by-graphs} shows the results.  
We can see the BA with finetuning is close to that without finetuing in all $\rho$ and when the trigger node size is not large (i.e., $n_{tri}=3,4$). This shows the backdoor effect is persistent. 

\vspace{+0.5mm}
\noindent {\bf (Stealthiness) Similarity between the backdoored graphs and clean graphs is large:} 
We further quantitatively compare the structure similarity between the generated backdoored graphs and the clean graphs, where we use the metrics NetSim and DeltaCon proposed in~\cite{wills2020metrics}. 
Table \ref{table:metric} shows the similarity results over all training graphs. 
We observe the backdoored graphs and their clean counterparts are structurally close (except BITCOIN where one possible reason could be  the BITCOIN graph is very sparse). 

The above results imply that empirical defenses based on finetuning and structure similarity test are hard to detect or remove the backdoor trigger. Also, empirical defenses are always broken by advanced/adaptive attacks~\cite{zhang2021backdoor}. Hence, it is necessary to develop certified defenses for backdoored FedGL. More details see Section~\ref{sec:defense}.     

\begin{table}[!t]
\small
\renewcommand{\arraystretch}{1.0}
%\vspace{-2mm}
%\addtolength{\tabcolsep}{-0.5pt}
\addtolength{\tabcolsep}{+1pt}
\caption{Impact of different modules in our adaptive trigger generator on the (MA/BA) performance on Bitcoin. 
T-L: Trigger-Location; T-S: Trigger-Shape; E-V: Edge-View; Cus-T: Customized-Trigger; Def-T: Definable-Trigger.}
% \vspace{-4mm}
\centering
    \begin{tabular}{c||c c|c|c c}
    \toprule
    {\bf Models} & T-L & T-S & T-L w/o E-V & Cus-T & Def-T \\
    \hline
    \hline
    \rowcolor{gray!28}
    {\bf (a)} & \ding{51} & \ding{51} & & {\bf 0.72 / 0.99} & {\bf 0.72 / 0.99} \\
    % \hline
    {\bf (b)} &  & \ding{51} & \ding{51} & 0.72 / 0.83 & 0.72 / 0.84 \\
    {\bf (c)} &  & \ding{51} & & - & 0.71 / 0.81 \\
    %\hline
    {\bf (d)} & \ding{51} & & & 0.72 / 0.72 & 0.72 / 0.77 \\
    %\hline
    {\bf (e)} &  &  &  & - & 0.71 / 0.63 \\
     \bottomrule
    \end{tabular}
\label{table:ablation}  
% \vspace{-2mm}
\end{table}

\vspace{-2mm}
\subsubsection{Ablation study}
In this experiment, we examine 
the contribution of each module in our trigger generator. 
The modules include Trigger-Location (based on the Edge-View module), Trigger-Shape, Customized-Trigger, and Definable-Trigger.  
For simplicity, we test on BITCOIN, and the other datasets show similar observations. 
The results are summarized in Table \ref{table:ablation} with  $\rho = 20\%$, and $n_{tri}=4$ in Definable-Trigger and $n^*_{tri}=5$ in Customized-Trigger. 

{\bf (a)} The whole generator as a reference. 
{\bf (b)} We exclude only the Edge-View sub-module in Trigger-Location, indicating that trigger locations in  graphs are computed solely based on the rank of the node features. 
{\bf (c)} We remove  Trigger-Location and decide the trigger location in each graph randomly. 
Compared with {\bf (a)}, the significant BA reductions of 16\% for Customized-Trigger, and 15\% and 18\% for Definable-Trigger in {\bf (b)} and {\bf (c)} demonstrate that the Trigger-Location module excels at selecting important nodes 
in the graphs. 
{\bf (d)} We remove  Trigger-Shape and use the ER model~\cite{gilbert1959random} to decide the trigger shape. The reductions of 27\% and 22\% in BA underscore the superior effectiveness of our Trigger-Shape module in learning trigger shapes. {\bf (e)} We remove both the Trigger-Shape and Trigger-Location modules, resorting to a random method for trigger location selection and an ER model for trigger shape determination. 
The substantial 36\% reduction in BA proves the strong competition of our method for backdoor attacks on FedGL.

\begin{figure}[!t]
% \vspace{-2mm}
\centering	
\includegraphics[width=0.48\textwidth]%0.48
{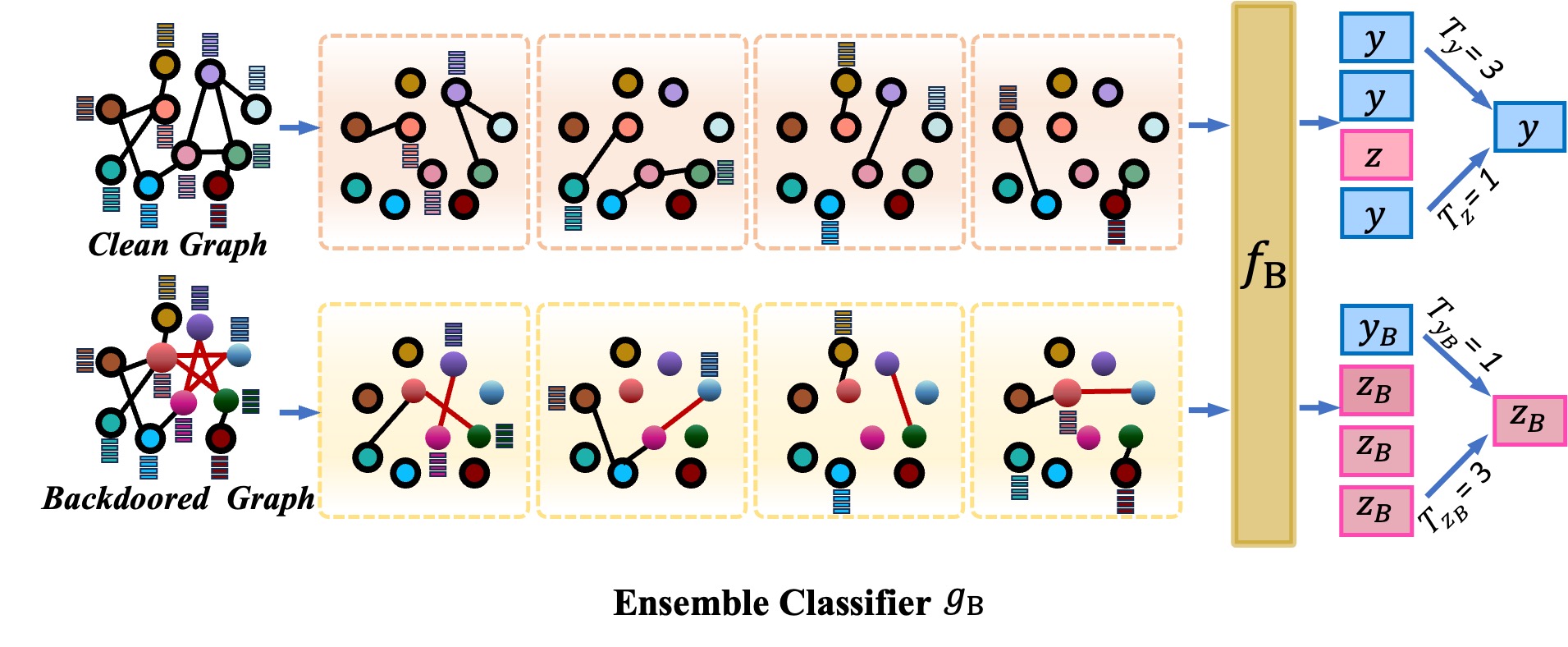} 
\vspace{-8mm}
\caption{Overview of our proposed certified defense.}
\label{fig:overview-defense}
%\vspace{-2mm}
\end{figure}

%\vspace{-8mm}
\section{Certified Defense for 
FedGL}
\label{sec:defense}
% \vspace{-2mm}

In this section, we design certified defenses for (backdoored) FedGL. 
Suppose we have learnt a backdoored FedGL for graph classification. 
We aim to build a certifiably robust defense mechanism such that 
the graph classifier: 1)  
provably predicts the correct label for clean testing graphs injected with arbitrary trigger  
with a bounded size; and 2) provably predicts a non-target label for backdoored testing graphs.  
This has the implication that benign clients' performance are provably kept, 
while the malicious clients' backdoored effect are provably removed during testing.    
 
Our defense includes three key steps: 1) dividing a (clean or backdoored) testing graph into multiple subgraphs; 2) building a majority vote-based ensemble graph classifier on the subgraphs; and 3)  deriving the robustness guarantees of the ensemble graph classifier against arbitrary trigger. 
Figure~\ref{fig:overview-defense} overviews our defense. 

\subsection{Graph Division into Subgraphs}
%\vspace{-2mm}

Recall that a backdoor attack can arbitrarily perturb the edges $\mathcal{E}$ and node features ${\bf X}$ in a graph $G=(\mathcal{V},\mathcal{E},{\bf X})$ as the trigger can be put in any location with any shape. 
To defend against this attack, our main idea is to design a {deterministic} 
function $h$\footnote{We emphasize that the function should be independent of the graph structure and node features. Otherwise, an attacker may possibly ``reverse engineer'' the function to find the relation between the output and input.} to divide $G$ into different subgraphs, such that each edge and node (feature) in $G$ is deterministically mapped into only one subgraph. 

\vspace{+0.5mm}
\noindent {\bf Hash function as the mapping:} 
We use the cryptographic hash function (e.g., MD5) as our mapping function. It takes input as a bit string and outputs an integer (e.g., 128-bit long with the integer range $[0,2^{128}-1]$). Here, we propose to use node indexes and stringify them as the input to the hash function. For instance, for a node $v$, we denote its string as $\textrm{str}(v)$.  
Given the $\textrm{str}(v)$ of every node $v \in \mathcal{V}$, we propose mapping the nodes and edges using the hash function $h$ and dividing a graph $G$ into multiple (e.g., $T$) subgraphs.

\vspace{+0.5mm}
\noindent {\bf Node feature and edge division:}
First, we divide node features into $T$ groups using the hash function $h$. Specifically, we compute the hash value
$\big( h[\mathrm{str}(u)] \, \mathrm{mod} \, T + 1 \big)$ for every node $u \in \mathcal{V}$, where $\mathrm{mod}$ is the modulo function. 
We use 
$\mathcal{V}^t$ to denote the set of nodes whose group index is $t$, i.e., $\mathcal{V}^t = \{u \in \mathcal{V}\ |\ h[\textrm{str}(u)]= t \}$, $t=1,2,\cdots T$. 
Correspondingly, we use ${\bf X}^t \in \mathbb{R}^{|\mathcal{V}| \times d}$ to denote the node features in the $t$th group. 
{With such grouping, we observe that nodes not in the $t$th group do not have features in ${\bf X}^t$}. To mitigate this, we simply set features of these nodes to be zeros. 
I.e., ${\bf X}_v^t = {\bf X}_v$, if $v \in \mathcal{V}^t$; and ${\bf X}_v^t = {\bf 0}$, otherwise.    

Next, we divide edges into $T$ groups. Specifically, we compute $h[\mathrm{str}(u) + \mathrm{str}(v)] \, \mathrm{mod} \, T+1$ 
for every edge $(u,v) \in \mathcal{E}, u\leq v$, where ``+'' means the string concatenation. 
We then let $h[\mathrm{str}(v) + \mathrm{str}(u)]=h[\mathrm{str}(u) + \mathrm{str}(v)]$. 
This is to ensure an undirected edge has a same hash value. 
We use 
$\mathcal{E}^t$ to denote the set of edges whose group index is $t$, i.e., $\mathcal{E}^t = \{(u,v) \in \mathcal{E}\ |\ h[\textrm{str}(u) + \textrm{str}(v)]= t \}$. 

Then, we construct $T$ subgraphs, i.e., ${G}^t = (\mathcal{V}, \mathcal{E}^t, {\bf X}^t)$ with $t=1,2,\cdots, T$, for a graph $G$. Notice the node features and edges are non-overlapped between different subgraphs. That is, ${\bf X}^i \cap {\bf X}^j = \emptyset , \mathcal{E}^i \cap \mathcal{E}^j =  \emptyset, \forall i,j \in \{1, 2, \cdots, T\}, i \neq j$. 
This is a requirement to enable deriving our robustness guarantee. 
{Note also that the subgraph does not need to be connected, as a graph classifier can still predict a label for a graph with  disconnected components.}

% \vspace{-1mm}
\subsection{Majority Vote-based Ensemble Classifier}
\label{sec:ensembleclassifier}
%\vspace{-1mm}

Let the backdoored graph classifier be  $f_B$.  
Given a clean testing graph $G$ (with true label $y$), we construct $T$ subgraphs $\{G^t\}$ using our graph division strategy and introduce a majority vote-based ensemble graph classifier $g_B$ to classify these $T$ subgraphs.  
Specifically, we denote by $T_l$ the number of subgraphs classified as the label $l$ by $f_B$, i.e., $T_l = {\textstyle \sum_{t=1}^T} \mathbbm{1} (f_B(G^t) = l)$. Then, we define 
$g_B$ as:
\begin{equation}
    g_B(G) = 
    {\arg\max}_{l \in \mathcal{Y}} \, T_l,
\label{equation: clean-ensemble}
\end{equation}
which returns a smaller 
index when ties exist. 
Let $y = g_B(G)$ by assuming the ensemble classifier accurately predicts the clean graph.  

Similarly, for a backdoored testing graph $G_B$  
(with the target label $y_B$), we construct $T$ subgraphs $\{G_B^t\}$ using the graph division strategy and denote by $T_{l_B}$ the number of subgraphs classified as the label $l_B$ by $f_B$, 
i.e., $T_{l_B} = {\textstyle \sum_{t=1}^T} \mathbbm{1} (f_B(G_B^t) = l_B)$. Then, we have: 
\begin{equation}
    g_B(G_B) = 
    {\arg\max}_{l_B \in \mathcal{Y}} \, T_{l_B}. 
\label{equation: backdoor-ensemble}
\end{equation}
Likewise, 
we let $y_B = g_B(G_B)$ by assuming the backdoored testing graph successfully triggers the backdoor.   

%\vspace{-2mm}
\subsection{Certified Robustness Guarantees}
\label{sec:certifiedrobustness}
%\vspace{-1mm}

With our graph division strategy and ensemble classifier, we can derive the robustness guarantee for clean graphs against backdoor trigger and backdoored graphs. 
{\bf \emph{Proofs 
are  deferred to Appendix~\ref{app:thmproofs}.}}

% \vspace{-1mm}
\subsubsection{Certified robustness w.r.t. clean graph}
Assume we have a backdoored graph $\tilde{G}$ generated from the clean graph $G$. 
We use $\tilde{G}^1, \tilde{G}^2, \cdots, \tilde{G}^T$ to denote the $T$ subgraphs from $\tilde{G}$ via the graph division strategy. Moreover, we denote by $\tilde{T}_l = \sum_{t=1}^T\mathbbm{1}(f_B(\tilde{G}^t)=l), \forall l \in \mathcal{Y}$, and $g_B(\tilde{G}) = {\arg\max}_{l \in \mathcal{Y}} \, \tilde{T}_l$. 
We aim to ensure $g_B(G) = g_B(\tilde{G})$ when the perturbation size  
induced by the backdoor trigger is bounded by a threshold (call \emph{certified perturbation size}), where 
the perturbation size is defined as the sum of the perturbed number of nodes 
(whose features can be arbitrarily modified) and 
edges w.r.t. $G$. 
Formally, we state the  theorem below: 

\begin{theorem}[Certified perturbation size w.r.t. clean graph]
% \vspace{-1mm}
Given a backdoored graph classifier $f_B$ and our ensemble graph classifier $g_B$. Given a clean testing graph $G$ with a label $y$ and its $T$ subgraphs $\{G^t\}_{t=1}^T$ produced by our graph division strategy. 
Suppose $T_y$ and $T_z$ are the largest and second largest frequency outputted by $f_B$ on predicting $\{G^t\}_{t=1}^T$. 
Let $m$ be the perturbation size induced by an arbitrary backdoor trigger and the respective backdoored graph is $\tilde{G}$.  
Then $g_B(G) = g_B(\tilde{G}) = y$, when $m$ satisfies: 
\begin{equation}
m \le  m^* =   
\lfloor \frac{T_y - T_z + \mathbbm{1}(y<z) -1}{2} \rfloor. 
\label{eq: T_y - T_z}
\end{equation}
\label{thm:certiclean}
%\vspace{-1mm}
\end{theorem}

We have below remarks of the theoretical result from Theorem~\ref{thm:certiclean}:
\begin{itemize}[leftmargin=*]

\item It can be applied for \emph{any} backdoored FedGL model. 

\item It holds for \emph{any} backdoored attack with a trigger that arbitrarily perturbs 
$m^*$ edges and nodes. 

\item It does not restrict the trigger to be connected. 

\item 
The robustness guarantee is true with a probability 100\%. 

\end{itemize}

Next, we further show 
our certified robustness guarantee is tight.

\begin{theorem}[Tightness of $m^*$]
% \vspace{-1mm}
For any $m$ satisfying $m > m^*$, 
there exists a base classifier $f'_B\neq f_B$ that will make $g_B$ misclassify $\tilde{G}$. 
That 
being said, 
it is impossible to derive a larger certified perturbation size than $m^*$ in Theorem~\ref{thm:certiclean}, without using extra information on 
$f_B$.  
\label{thm:tightclean}
% \vspace{-2mm}
\end{theorem}

\subsubsection{Certified robustness w.r.t. backdoored graphs} 
For a backdoored graph $G_B$, we consider its robustness against our defense strategy. 
We have the below theorem. 

\begin{theorem}[Certified (non-)backdoored graph] 
Let $f_B$ and $g_B$ be defined as Theorem~\ref{thm:certiclean}. 
Given a backdoored testing graph $G_B$ with a target label $y_B$ and its $T$ subgraphs $\{G_B^t\}_{t=1}^T$ produced by our graph division. 
Let $T_{y_B}$ and $T_{z_B}$ be the largest and second largest frequency outputted by $f_B$ on predicting $\{G_B^t\}_{t=1}^T$. 
Then if $T_{y_B} > T_{z_B} - \mathbbm{1}(y_B < z_B)$,  
$G_B$ is a certified backdoored graph for our defense, otherwise it is a certified non-backdoored graph. 

\label{thm:certibackdoor}
% \vspace{-1mm}
\end{theorem}
A certified backdoored graph means it provably evades our defense, while a certified non-backdoored graph means our defense provably predicts it as a non-target label.

\begin{figure*}[!t]
\centering	
\includegraphics[width=0.85\textwidth]{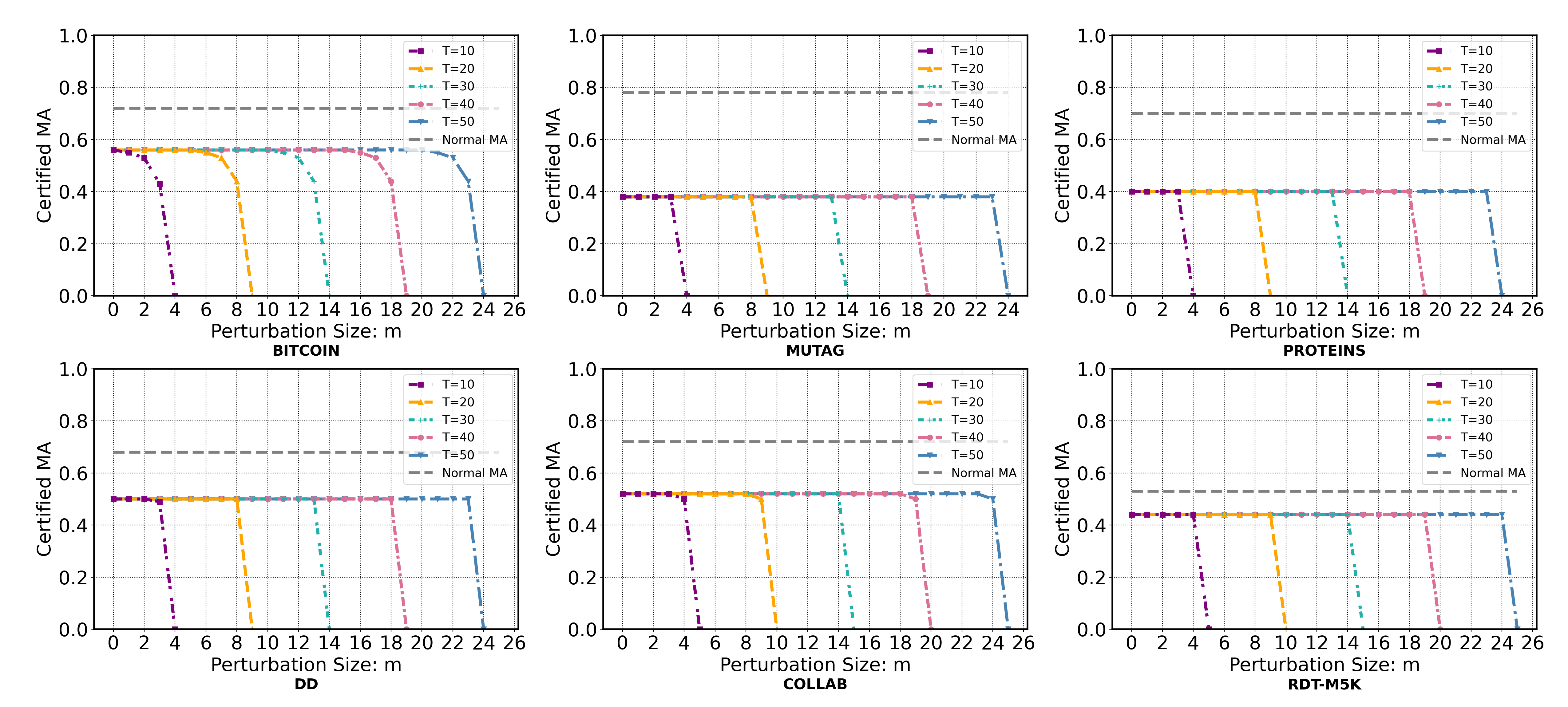}
\vspace{-3mm}
\caption{Certified MA vs. $T$. $100$ testing graphs are randomly sampled (and all testing graphs in MUTAG) for evaluation. Normal MA (under no attack and defense) is also reported for reference.} 
\label{fig:plot-defense}
\vspace{-4mm}
\end{figure*}

\begin{figure*}[!t]
\centering	
\includegraphics[width=0.85\textwidth]{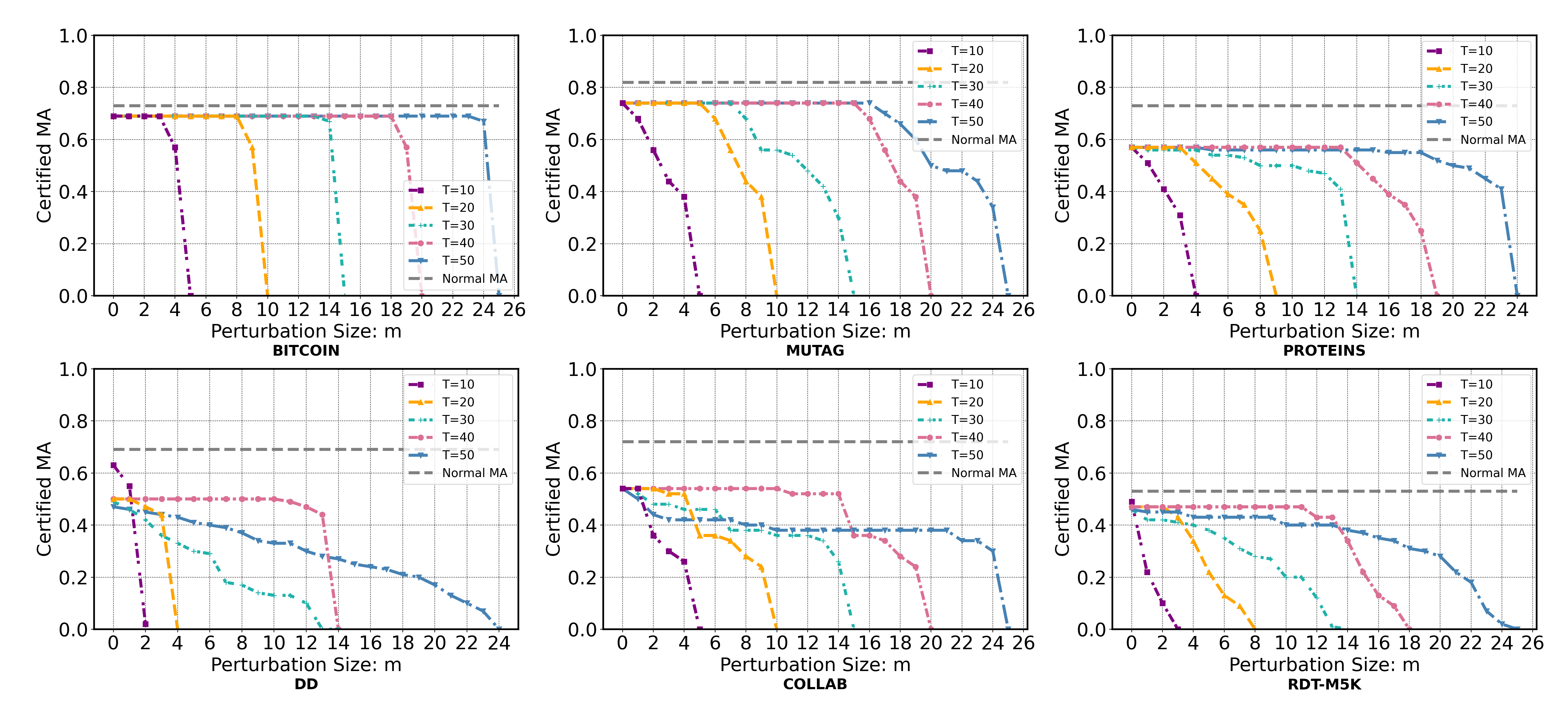}
\vspace{-2mm}
\caption{Certified MA vs. $T$, where we finetune the backdoored FedGL model with augmented subgraphs that are generated from the benign clients' training graphs. 
}
\label{fig:plot-defense-aug}
\vspace{-2mm}
\end{figure*}

\section{Certified Defense Results}
\label{sec:defenseresults}

\subsection{Experimental Setup}

\vspace{+0.5mm}
\noindent {\bf Parameter setup:} 
We first train the backdoored FedGL model under our Opt-DGBA (with a specified $\rho$, $n_{tri}$ in Definable-Trigger or $n^*_{tri}$ in Customized-Trigger). 
The trained backdoored graph classifier 
$f_B$ is then for  
both normal testing (i.e., on cleaning testing graphs) and backdoor testing (i.e., on the backdoored testing graphs generated by our Opt-GBA). Here, we only select successfully backdoored testing graphs for evaluation. 
Unless otherwise specified, we use $\rho = 20\%$, $n_{tri} = 4$, or $n^*_{tri} = 5$. 

To apply our defense, for each (clean/backdoored) testing graph, we use our graph division strategy to divide it into $T$ subgraphs and majority vote-based ensemble classifier $g_B$ to predict these subgraphs.  
The key hyperparameter in our defense is the number of subgraphs $T$. By default we set $T=30$. We also test its impact on the defense performance. 

\vspace{+0.5mm}
\noindent {\bf Evaluation metrics:} We use the certified accuracy~\cite{bojchevski2020efficient,wang2021certified,lai2023towards,scholten2022randomized} on the testing graphs for evaluation. 
\begin{itemize}[leftmargin=*]
\item {\bf Certified MA at perturbation size $m$:}  the fraction of the clean testing graphs that are provably classified as the true label against an arbitrary trigger whose size is $m$. 

\item {\bf Certified BA:} the fraction of the backdoored testing graphs that are provably classified as the target label against our defense. 
 
\end{itemize}

\begin{figure*}[!t]
\centering	
\includegraphics[width=0.85\textwidth]{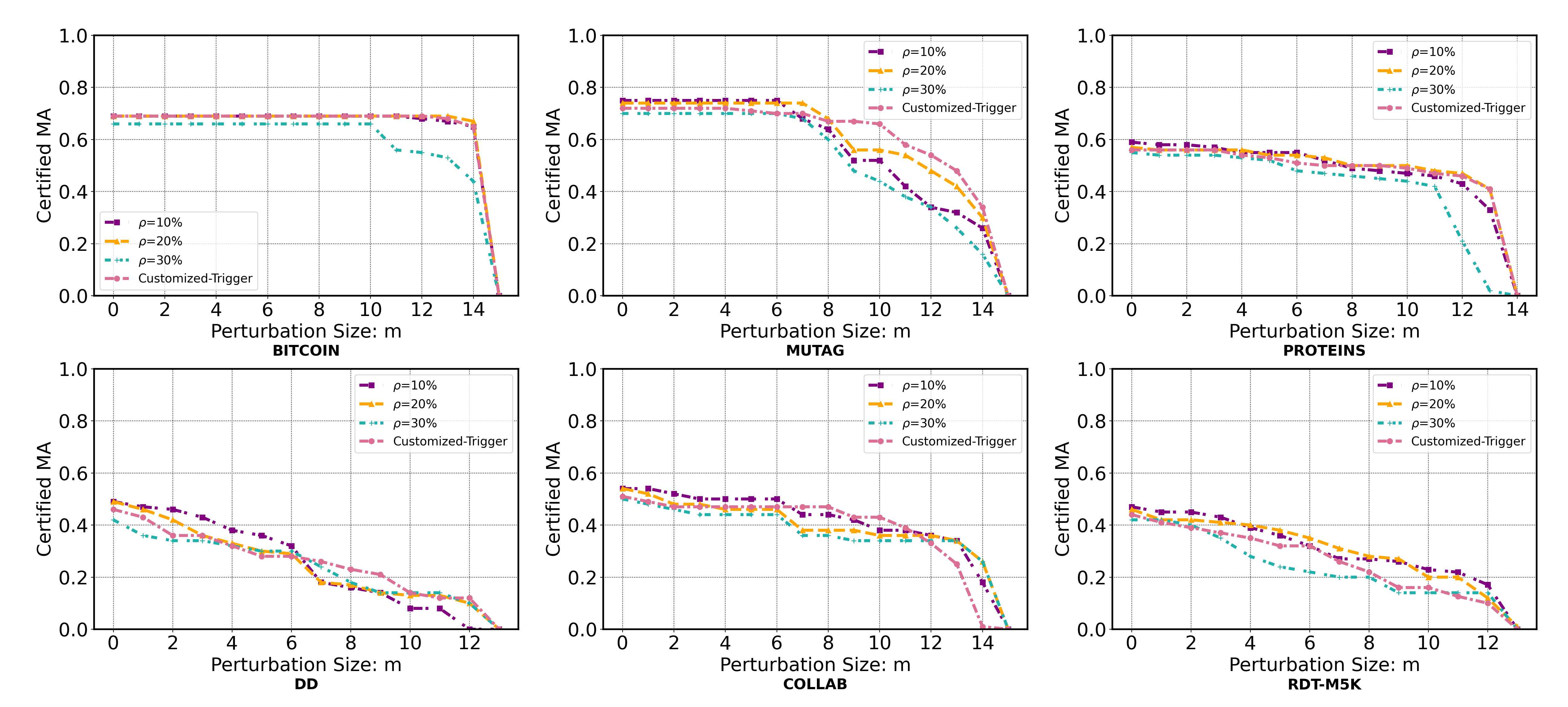}
\vspace{-2mm}
\caption{Certified MA vs. ratio $\rho$ of malicious clients ($T=30$, $n_{tri}=4$) 
}
\label{fig:plot-defense-aug-poison-2}
\vspace{-4mm}
\end{figure*}

\begin{figure*}[!t]
\centering	
\includegraphics[width=0.85\textwidth]{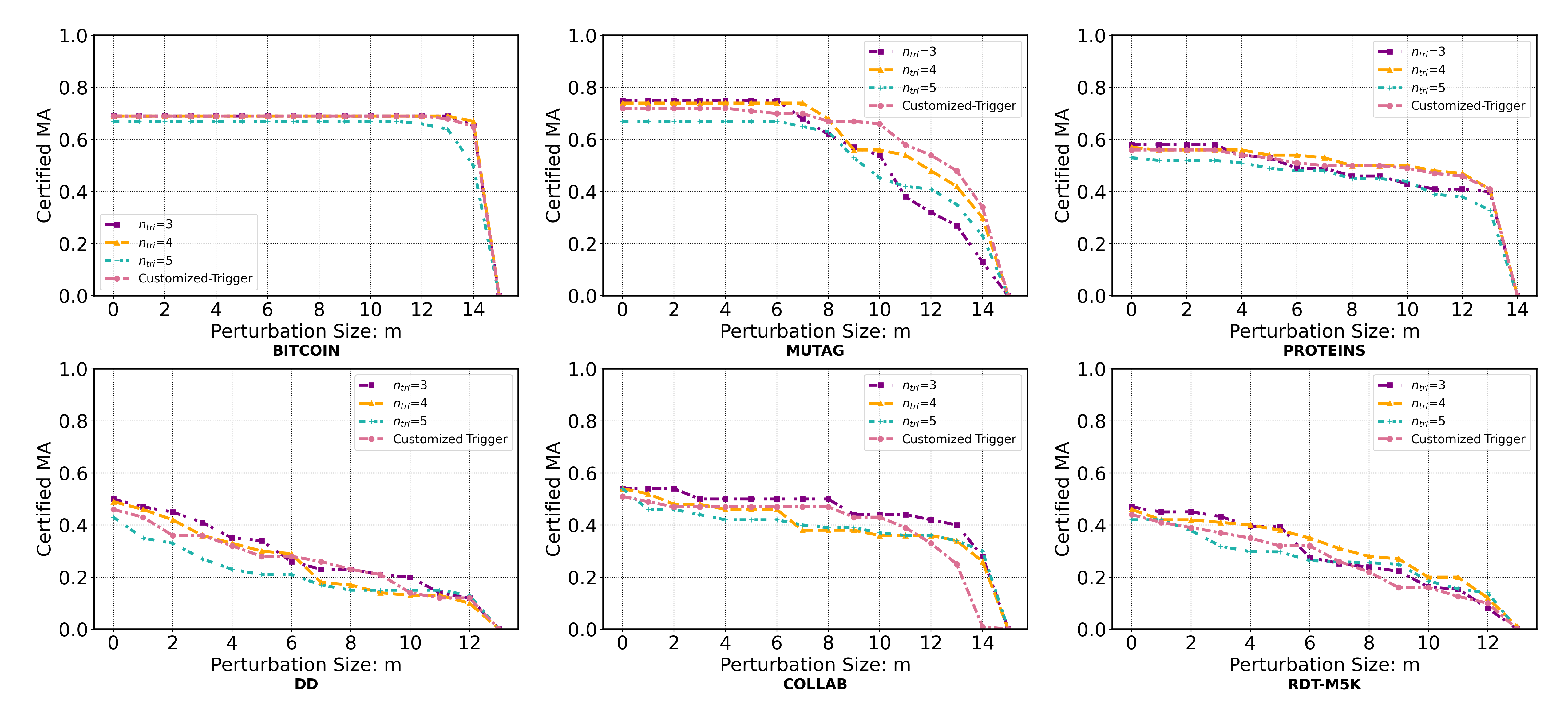}
\vspace{-2mm}
\caption{Certified MA vs. trigger node size $n_{tri}$ ($T=30$, $\rho=20\%$).  
}
\label{fig:plot-defense-aug-trigger-2}
\vspace{-2mm}
\end{figure*}

% \vspace{-2mm}
\subsection{Experimental Results}
%\vspace{-2mm}

\subsubsection{Results on certified MA}
In this experiment we show the results on certified MA against the backdoored FedGL model trained under our Opt-GDBA.  
{\bf \emph{More results are shown in Appendix~\ref{app:defenseresults}.}} 

\vspace{+0.5mm}
\noindent {\bf Certified MA vs. $T$:} Figure~\ref{fig:plot-defense} shows the certified MA at perturbation size $m$ vs. different $T$, where we test on all 63 testing graphs in MUTAG and randomly sampled 100 testing graphs in the other datasets. For reference, we also report the normal MA without our defense. 
In general, we observe our defense is  provably more robust (i.e., larger certified MA with larger certified perturbation size) when $T$ is larger. For instance, on BITCOIN, when $T=30,50$, 
certified MA are 44\% and 58\% with $m=13$,  
and the maximum certified perturbation size are $m^* = 13$ and 23, respectively. 
The reason is our majority vote-based ensemble  classifier could tolerate more perturbed subgraphs when $T$ increases. As an arbitrary node/edge perturbation induced by the trigger can affect at most one clean subgraph's prediction, a larger $T$ implies being robust to a larger $m$.  

Though effective, we still see a large gap between certified MA and normal MA on datasets such as MUATG and PROTEINS. 
This is due to the number of accurately predicted subgraphs (i.e, $T_y$ in Equation (\ref{eq: T_y - T_z})) after graph division is not  large enough on these datasets. 
We note that the backdoored FedGL training only uses the whole training graph. To enhance the certified MA, we propose to finetune the backdoored FedGL model with extra subgraphs created from benign clients' training graphs. Specifically, in each benign client, we use $T=\{10,20,...,50\}$ to generate a set of subgraphs for each training graph and pick one subgraph from each $T$. These subgraphs have the same label as the raw graph.  
Figure~\ref{fig:plot-defense-aug} shows the results. 
We observe the finetuned model with clean subgraphs yield a significantly higher certified MA on the relatively sparser/smaller datasets (e.g., BITCOIN, MUTAG, and PROTEINS). This implies the finetuned model learns a correct mapping between the subgraphs and the true label, and hence improves the accuracy of subgraphs created from the testing graphs. 
In contrast, we see a drop of certified MA on relatively denser/larger datasets (e.g., DD, COLLAB, and RDT-M5K). This is possibly because the finetuned model is hard to associate  
both the large dense graphs and their generated much smaller and sparser subgraphs with the true label. 

The above results suggest that, in practice, to enhance the certified MA of a backdoored FedGL model, we could augment the training graphs with their   subgraphs on small/sparse datasets, but may not on large/denser datasets.  

\vspace{+0.5mm}
\noindent {\bf Certified MA vs. $\rho$:} 
In this experiment, we assess the defense performance on Opt-GDBA attacked FedGL models with varying $\rho$ of malicious clients. 
Figure~\ref{fig:plot-defense-aug-poison-2} shows the results with $T=30$ and $n_{tri}=4$. We observe the certified MA and certified perturbation size are similar  with different $\rho$'s (except a slight drop when $\rho=30\%$). 
This is primarily because the Opt-GDBA's MA and BA are relatively stable across different $\rho$, as shown in Table~\ref{table:GraphDBA-results}. 
This ensures the number of correctly predicted subgraphs via our ensemble classifier is also close in different $\rho$'s, and so do the certified MA and certified perturbation size.  

\vspace{+0.5mm}
\noindent {\bf Certified MA vs. $n_{tri}$:} This experiment explores the defense performance with varying trigger node sizes used by Opt-GDBA. Figure~\ref{fig:plot-defense-aug-trigger-2} shows the results. 
Similarly, 
our defense achieves similar certified MA in general, with large $n_{tri}$ slightly reduces the certified MA. 

% \vspace{-2mm}
\subsubsection{Results on certified BA} 
In this experiment, we evaluate the robustness of the backdoored testing graphs generated by our Opt-GDBA under our defense.   
Table \ref{table:certified-BA} shows  the results of 
certified BA.   
We observe the certified BA is 0 
in all $T$, $\rho$, and $n_{tri}$. 
Recall our graph division strategy ensures the divided subgraphs have non-overlapping edges and node features. 
The above results can be attributed to two aspects: 
the trigger in the backdoored testing graph is  separated into: 1) a few subgraphs that are still classified as the \emph{target label}, but the other majority subgraphs are mostly classified as a non-target label (actually the true label in most cases); 
or 2) a large number of the subgraphs that makes it difficult to form any effective trigger in the subgraphs.  
In either case, the number of successful backdoored subgraphs is a  minority. Hence, with the majority voting, all the backdoored testing graphs are misclassified as a \emph{non-target label}.   
The results imply the backdoored testing graphs generated by our Opt-GDBA are completely 
broken by our graph division + ensemble classifier based defense. 

We also calculate MA on the generated backdoored graphs (which have correct predictions without backdoor) under our defense.  
We obtain $\geq 92\%$ MA in all datasets, where 4 datasets are 100\%. \emph{This means our defense does not/marginally affect clean labels, so the FedGL's utility is still maintained.}  This is because the proposed defense is mainly designed to affect the backdoored effect in the backdoored subgraphs, but not affect the utility of clean subgraphs. 

\begin{table}[!t]
\small
%\scriptsize
\renewcommand{\arraystretch}{1.0}
\addtolength{\tabcolsep}{-4pt}
\caption{Certified BA and MA of backdoored graphs under our defense (in all $T$, $\rho$, and $n_{tri}$).}
\centering
    \begin{tabular}{c||c c c c c c }
    \toprule
    {\bf Datasets} & BITCOIN & MUTAG & PROTEINS & DD & COLLAB & RDT \\ 
    \hline
    \hline
    Certified BA & 0 & 0  & 0  & 0  & 0  & 0  \\
    % \hline
    MA & 1.00 & 1.00  & 1.00  & 1.00 & 0.96   & 0.92 \\
    \bottomrule
    \end{tabular}
\label{table:certified-BA}  
\end{table}

% \vspace{-2mm}
\section{Related Work}
\label{sec:related}

%\vspace{+0.5mm}
\noindent{\bf Backdoor attacks on centralized learning for non-graph data and defenses:} 
Extensive works have shown centralized machine learning models for non-graph data, such as image~\cite{gu2017badnets,chen2017targeted,liu2017trojaning,li2018hu,clements2018hardware,tran2018spectral,yao2019latent,wenger2021backdoor,salem2022dynamic,tao2023distribution}, text~\cite{gan2022triggerless,pan2022hidden,qi2021hidden,chen2021badpre,codebackdoor}, audio~\cite{roy2017backdoor,shi2022audio,ge2023data,guo2023masterkey}, and video~\cite{zhao2020clean,XieYH23}, are vulnerable to backdoor attacks. 
A backdoored model produces attacker-desired behaviors when the same trigger is injected into testing data. 
Gu et al.~\cite{gu2017badnets} proposed the first backdoor attack, called BadNet, 
on image classifiers. The attack injects a trigger (e.g., a sticker with yellow square) into ``STOP'' sign from the U.S. stop signs database and changes their labels to the ``SPEED'' sign. The trained backdoored image classifier then predicts a ``STOP'' sign with the same sticker trigger to be the ``SPEED'' sign.  

Many empirical defenses~\cite{chen2017targeted,liu2017trojaning,liu2018fine,wang2019neural,gao2019strip,liu2019abs,guo2019tabor,wang2020practical} have been proposed to mitigate backdoor attacks. For instance, 
Wang et al.~\cite{wang2019neural} proposed Neural Cleanse to detect and reverse engineer the trigger. However, all these defenses are broken by adaptive attacks~\cite{weber2023rab}. 
These two works~\cite{wang2020certifying,weber2023rab} proposed provable defenses against backdoor attacks in the image domain. However, they are shown to have insufficient effectiveness against backdoor attacks. In addition, they cannot be applied to inputs with different sizes.  

%\vspace{+0.5mm}
\noindent{\bf Backdoor attacks on federated learning for non-graph data and defenses:} 
 Backdoor attacks on FL are categorized as centralized backdoor attack (CBA)~\cite{bhagoji2019analyzing, gong2022backdoor, wang2020attack, zhang2022neurotoxin}, where all malicious clients use a shared trigger, and distributed backdoor attack (DBA)~\cite{xie2019dba}, where each malicious client uses its own defined trigger. 
For instance, Bagdasaryan et al.~\cite{bagdasaryan2020backdoor} designed the first CBA via model replacement. 
Inspired by the distributed learning property of FL,  Xie et al.~\cite{xie2019dba} designed the first DBA which is shown to be more persistent and stealthy than CBAs on FL.  

Many empirical defenses~\cite{sun2019can,ozdayi2021defending,rieger2022deepsight,nguyenflame,zhang2022fldetector,cao2023fedrecover} have been proposed, and can be performed in the FL stage of pre-aggregation~\cite{rieger2022deepsight,nguyenflame}, in-aggregation~\cite{sun2019can,ozdayi2021defending,zhang2022fldetector}, and post-aggregation~\cite{cao2023fedrecover}. 
However, they can only defend against known attack techniques, and an adversary aware of the existence of these defenses can break them~\cite{wang2020attack}.
Existing certified defenses~\cite{cao2021provably,cao2022flcert,xie2021crfl,xie2023unraveling} for FL can only tolerate a very small number of malicious clients and/or incur a large computation/communication cost for clients.

\vspace{+0.5mm}
\noindent{\bf Backdoor attacks on centralized graph learning and defenses:}
Unlike non-graph data that can be represented via Cartesian coordinates and have fixed input size, graphs cannot do so and different graphs often have varying sizes, making the trigger hard to be defined. To address this, two recent works~\cite{zhang2021backdoor,xi2021graph} propose to use \emph{subgraph} as a trigger. Zhang et al.~\cite{zhang2021backdoor} use a random subgraph as the trigger shape, which is generated by random graph generation models (such as the Erdős-Rényi~\cite{gilbert1959random}, Small World~\cite{watts1998collective}, and Preferential Attachment~\cite{barabasi1999emergence}), and pick random nodes as the trigger location. 
Instead of using a random trigger shape, Xi et al. \cite{xi2021graph} designed a trigger generator to learn to generate the trigger shape for each 
graph using its edge and node feature information. However, the trigger randomly chooses nodes 
as the trigger location. 

Zhang et al.~\cite{zhang2021backdoor} proposed a certified defense for a backdoored graph classifier by extending randomized ablation~\cite{levine2020robustness} for image classifiers. 
Specifically, they built a randomized subgraph sampling based defense mechanism to ensure the backdoored graph classifier provably predicts the same label for a testing graph if the injected trigger has a size less than a threshold.  
However, their defense is limited to edge perturbation and their robustness guarantee could be incorrect with a certain probability.

\vspace{+0.5mm}
\noindent{\bf Backdoor attacks on federated graph learning:} 
Xu et al.~\cite{xu2022more} is the only work studying backdoor attacks on FedFL. It is inspired by \cite{zhang2021backdoor,xie2019dba} with random subgraph as a trigger. We showed their backdoor performance is not good enough with a smaller trigger.  

\vspace{+0.5mm}
\noindent {\bf Majority-voting based ensemble for certified defenses:} 
The key insight of this type of defense is to ensure only a bounded number of corrupted votes/predictions (each prediction is treated as a vote) are changed with a bounded adversarial perturbation. This idea has been used in certified defenses against adversarial patch attacks~\cite{levine2020randomized,xiang2021patchguard}, data poisoning attacks~\cite{jia2021intrinsic,levine2020deep,jia2022certified}, and  others~\cite{hong2022unicr,pei2023textguard,zhang2024text}. 
The key difference among these methods is that they create problem-dependent voters for the majority vote.
A closely relevant work to ours is \cite{yanggnncert}, but 
they have two key differences.  
First, the studied problem is different. \cite{yanggnncert} proposes a majority-voting strategy for GL to defend against {evasion} attacks, while ours for FedGL to defend against distributed backdoor attacks.
Second, the graph division strategy is different. \cite{yanggnncert} divides a graph into \emph{overlapped} subgraphs, which facilitates deriving the robustness guarantee against graph perturbation \emph{or} node (feature) perturbation, but not both. Instead, our method can concurrently defend against both graph \emph{and} node (feature) perturbations.

%\vspace{-2mm}
\section{Conclusion}
\label{sec:conlcusion}

We study the robustness of FedGL 
from both the attacker's and defender's perspective. 
We first design an effective, stealthy, and persistent DBA on FedGL. 
Instead of using a random (centralized or distributed) trigger that is injected into random position in a graph, our attack develops a trigger generator that adaptively learns the important trigger location and shape per backdoored graph. Our attack results show 
existing empirical defenses based on backdoor detection or removal are ineffective.  Then, we further develop a 
certified defense for backdoored FedGL model based on 
graph division and majority vote-based ensemble. We derive the certified robustness as well as its tightness w.r.t. clean graphs against arbitrary trigger and backdoored graphs generated by our attack.

\section*{Acknowledgments} 
We thank all anonymous reviewers for
the constructive comments. Li is
partially supported by the National Natural Science Foundation of China under Grant No. 62072208, Key Research and Development Projects of Jilin Province under Grant No. 20240302090GX. 
Hong is partially supported by the National Science
Foundation under grant Nos. CNS-2302689, CNS-2308730, CNS-2319277 and CMMI-2326341.
Wang is partially supported by the National Science Foundation under grant Nos. ECCS-2216926, CNS-2241713, CNS-2331302 and CNS-2339686. 
Any opinions, findings and conclusions or recommendations expressed in this material are those of the author(s) and do not necessarily reflect the views of the funding agencies.

\bibliographystyle{ACM-Reference-Format}
\bibliography{cite,alan}

\clearpage
\appendix

\section{Network Architecture in Opt-GDBA}
\label{app:netarch}

\begin{table}[!h]
%\vspace{-2mm}
\renewcommand{\arraystretch}{1}
\addtolength{\tabcolsep}{-4pt}
\caption{Detailed Network architecture of the adaptive trigger generator in Opt-GDBA. $G^i$ is a clean graph from client $i$, and $G_B^i$ is its backdoored counterpart. Trigger $\mathcal{V}_{tri}^i$ indicates $\mathcal{V}_{def}^i$ or $\mathcal{V}_{cus}^i$, and $n = |
\mathcal{V}^i|$. $d$ is the feature dimension. Mask is a binary matrix taking a value $1$ at  position $\mathcal{V}_{tri}^i$ and $0$ otherwise, signifying our focus solely on edges and node features at the trigger location.
}
\vspace{-2mm}
\centering
    {\scriptsize
    \begin{tabular}{c||c|c|c}
    \toprule
    {\bf Network} & {\bf Input} & {\bf Network Architecture} & {\bf Output} \\
     \hline
     \hline
      & & Linear-$n  $ \& ReLU \& Dropout-0.05 & \\
    $\mathrm{EdgeView} $ & ${\bf A}^i \in \mathbb{R}^{n \times n}$ & Linear-$n  $ \& ReLU \& Dropout-0.05 & ${\bf e}^i \in \mathbb{R}^{n}$ \\
     & & Linear-$n$ \& Sigmoid & \\ 
      & &  AvgPooling-$n$ & \\
     \hline
     & & Linear-$d  $ \& ReLU \& Dropout-0.05 & \\
     $\mathrm{NodeView} $ & ${\bf X}^i \in \mathbb{R}^{n \times d}$ & Linear-$d $ \& ReLU \& Dropout-0.05 & ${\bf n}^i \in \mathbb{R}^{n}$\\
     & & Linear-$d $ \& Sigmoid & \\
     & &  AvgPooling-$d$ & \\
     \hline
      & & Linear-$n  $ \& ReLU \& Dropout-0.05 & \\
      
    $\mathrm{EdgeAtt} $& ${\bf A}^i_B \in \mathbb{R}^{n\times n}$ & Linear-$n $ \& ReLU \& Dropout-0.05 & ${\bf E}^i_{tri} \in \mathbb{R}^{n_{tri} \times n_{tri}}$ \\
     & $\mathcal{V}_{tri}^i \in \mathbb{R}^{n_{tri} \times n_{tri}}$ & Linear-$n  $ \& Sigmoid  & \\
     & & Mask-$n_{tri}$ & \\
     \hline
     & & Linear-$d$ \& ReLU \& Dropout-0.05 & \\
    $\mathrm{NodeAtt} $ & ${\bf X}^i_B \in \mathbb{R}^{n \times d}$ & Linear-$d$ \& ReLU \& Dropout-0.05 & ${\bf N}^i_{tri} \in \mathbb{R}^{n_{tri} \times d}$\\
     & $\mathcal{V}_{tri}^i \in \mathbb{R}^{n_{tri} \times n_{tri}}$& Linear-$d $ \& ReLU  &  \\
      & & Mask-$d$ & \\
    \hline
    $\mathrm{EdgeEmb} $& %$i \in 1 \times 1$ 
    & Linear-$n \times n$ & ${\bf I}_e \in \mathbb{R}^{n_{tri} \times n_{tri}}$ \\
    &$\mathcal{V}_{tri}^i \in \mathbb{R}^{n_{tri} \times n_{tri}}$ & Mask-$n_{tri}$ & \\
     \hline
    $\mathrm{NodeEmb} $& 
    % $i \in 1 \times 1$ 
    & Linear-$n \times d$ & ${\bf I}_n \in \mathbb{R}^{n_{tri} \times d}$ \\
    & $\mathcal{V}_{tri}^i \in \mathbb{R}^{n_{tri} \times n_{tri}}$& Mask-$d$ & \\
    \bottomrule
    \end{tabular}}
\label{table:net}  
%\vspace{-4mm}
\end{table}

\begin{table*}[!t]
\footnotesize
%\small
\renewcommand{\arraystretch}{0.9}
\addtolength{\tabcolsep}{0.2pt}
\caption{Dataset statistics and training/testing graphs.}
\vspace{-2mm}
\centering
    \begin{tabular}{c||c|c|c|c|c c c c c|c c c c c}
    \toprule
    Datasets & \multicolumn{1}{c}{$\#$Graphs} & \multicolumn{1}{c}{$\#$Classes} & \multicolumn{1}{c}{Avg. $\#\text{Node}$} & \multicolumn{1}{c}{Avg. $\#\text{
    Edge}$} & \multicolumn{5}{c}{$\#$Training graphs (per class)} & \multicolumn{5}{c}{$\#$Testing graphs (per class)} \\
    %\cline{6-15}
    & & & & & 1 &   2 &   3 &   4 & 5 & 1 &   2 &   3 & 4 & 5\\
    \hline
    \hline
     BITCOIN & 658 & 2 & 14.64 & 14.18 & 219 & 219 & - & - & -& 110 & 110 & - & - & - \\
    % \hline
    MUTAG & 188 & 2 & 17.93 & 19.80  & 83 & 42 & - & - & - & 42 & 21 & - & - & - \\
    % \hline
    PROTEINS & 1110 & 2 & 37.72 & 70.35  & 440 & 300 & - & -&- & 220 &  150 &  - & - & -\\
    % \hline
    DD & 950 & 2 & 208.3 & 518.76 & 330 & 303 & -& - & -& 165 & 152  &  - & - & -\\
    % \hline
  
    COLLAB & 4981 & 3 & 73.49 & 2336.66 & 517 & 1589 & 1215 & - & - & 258 & 794 & 608 & - & - \\
    % \hline
    RDT-M5K & 2613 & 5 & 208.0 & 227.66 &  120 & 556 & 423  & 545 & 98 & 61 &  277 & 212  &  272 & 49 \\
    \bottomrule
    \end{tabular}
\label{table:datasets}
\vspace{-2mm}
\end{table*}

%\vspace{-2mm}
\section{Proofs}
\label{app:thmproofs}

\noindent {\bf Proof of Theorem~\ref{thm:certiclean}:} Assume the perturbation size is $m$. Under our graph division strategy,  at most $m$ of the $T$ subgraphs in $G^1, G^2, \cdots, G^T$ are adversarially perturbed by the trigger. In other words, at most 
$m$ of the $T$ subgraphs in $\tilde{G}^1, \tilde{G}^2, \cdots, \tilde{G}^T$ are different from $G^1, G^2, \cdots, G^T$. Hence at most $m$ predicted labels  
on $\tilde{G}^1, \tilde{G}^2, \cdots, \tilde{G}^T$ are different from those on $G^1, G^2, \cdots, G^T$. 
Suppose $T_y$ and $T_z$ represent the largest and second large number outputted by $f_B$ on predicting $G^1, G^2, \cdots, G^T$, and  
$\tilde{T}_y$ and $\tilde{T}_z$ denote the corresponding numbers  
outputted by $f_B$ on predicting $\tilde{G}^1, \tilde{G}^2, \cdots, \tilde{G}^T$.  
 Then we have the following equations: 
\begin{equation}
\begin{aligned}
    & T_y-m \le \tilde{T}_y \le T_y+m, \\
    & T_z-m \le \tilde{T}_z \le T_z+m. 
\label{equation: T_y - {T}_y & T_z - {T}_z}
\end{aligned}
\end{equation}
To ensure our ensemble graph classifier $g_B$ still predicts $y$ for  
$G^p$, we require $\tilde{T}_y > (\tilde{T}_z - \mathbbm{1}(y<z))$, where $\mathbbm{1}(y<z)$ is due to the tie breaking mechanism (i.e., we  choose a label with a smaller number when ties exist). Combing with Equation (\ref{equation: T_y - {T}_y & T_z - {T}_z}), we require 
$$T_y-m > T_z+m - \mathbbm{1}(y<z),$$
Or $$m < \frac{T_y - T_z + \mathbbm{1}(y<z)}{2}.$$  
\vspace{+1mm}
\noindent {\bf Proof of Theorem~\ref{thm:tightclean}:} 
We use proof by contradiction. 
Specifically, we show that, 
for any $m$ satisfying $m > m^*$, i.e., at least $m^*+1$ perturbation size (or $m^*+1$ subgraphs $\{G^t\}_{t=1}^T$ are corrupted), there exists a base classifier $f'_B$
that makes the ensemble classifier $g_B$ do not 
%that is not guaranteed to 
predict $\tilde{G}$ as the label $y$, i.e., $g_B(\tilde{G}) \neq g_B({G}) = y$. 

For simplicity, we assume $m^*+1$ out of the $T$ graphs $\{G^t\}_{t=1}^T$  are corrupted.  
That is, we select $m^*+1$ sub-graphs among $\{G^t\}_{t=1}^T$ that are predicted as the label $y$ and corrupt them. Moreover, we let $f_B'$ predict the label $z$ for all the  $m^*+1$ corrupted sub-graphs, i.e., $z=\arg \max_{l\neq y}(T_l-\mathbbm{1}( y < l))$.  Hence, we have $$\tilde{T}_y=T_y-(m^*+1), \, \tilde{T}_z=T_z+(m^*+1).$$ 
Since $m^*$ is the largest number such that Equation \ref{equation: T_y - {T}_y & T_z - {T}_z} is satisfied, 
i.e., $T_y -m^* >   T_z - \mathbbm{1}(y<z) + m^*$ holds. Thus, 
$$T_y -(m^*+1) \leq   T_z - \mathbbm{1}(y<z) + (m^*+1).$$ 
Or $\tilde{T}_y \leq \tilde{T}_z$. Hence, under $f_B'$, $g_B(\tilde{G}) = z \neq y = g_B({G})$, if $y>z$. 

\vspace{+1mm}
\noindent {\bf Proof of Theorem~\ref{thm:certibackdoor}:}
Our ensemble graph classifier $g_B$ returns the label with the largest frequency of the predictions on $T$ subgrpah, or the label with a smaller index where there are ties.   
Hence, to ensure the backdoored graph $G_B$ is still predicted as the target label $y_B$ under $g_B$, it must satisfy $T_{y_B} > T_{z_B}$ or  $T_{y_B} =  T_{z_B}$ if $y_B < z_B$, or  $T_{y_B} > T_{z_B} - \mathbbm{1}(y_B < z_B)$. Otherwise, it will be provably predicted as a non-target label.

\section{Dataset Statistics and Descriptions}
\label{app:dataset}
%\vspace{-2mm}
Table~\ref{table:datasets} shows the dataset statistics and training/testing set. 
The descriptions of these graphs datasets are as below: 

%\vspace{+0.5mm}
\noindent {\bf BITCOIN:} 
It is used for graph-based fraudulent transaction detection. We obtain $658$ labeled transactions, where labels $0$ and $1$ correspond to illicit and licit transactions, respectively. 

\vspace{+0.5mm}
\noindent {\bf MUTAG:} 
It comprises of molecular data. Each molecule is associated with a binary label indicating its classification as either a steroid compound or not. 

\vspace{+0.5mm}
\noindent {\bf PROTEINS:} 
It is used for protein structure analysis. In this dataset, the nodes represent three primary elements of a protein, including helix, sheet, and turn; the edges signify the relationships between these structural elements.

\vspace{+0.5mm}
\noindent {\bf DD:} 
It includes protein structures, with each protein represented as a graph. The 
task is to classify these protein structures into either enzymes or non-enzymes.

\vspace{+0.5mm}
\noindent {\bf COLLAB:}
It is 
a collection of scientific collaborations. Labels $0$, $1$, and $2$ denote a researcher's ego network field, specifically High Energy Physics, Condensed Matter Physics, and Astro Physics. 
 
\vspace{+0.5mm}
\noindent {\bf RDT-M5K:} 
It is a social dataset. Each graph 
represents an online discussion thread, where nodes represent users and edges signify discussions between users. Additionally, each graph is labeled with the community or subreddit to which it belongs.

\begin{table*}[!t]
%\small
\footnotesize
\renewcommand{\arraystretch}{0.9}
\addtolength{\tabcolsep}{1pt}
\caption{Comprehensive results of the compared attacks. 
}
\vspace{-2mm}
\centering
    \begin{tabular}{c||c|c c c|c|c c|c c|c c}
    \toprule
    & \multicolumn{1}{c}{ } & \multicolumn{3}{c}{{\bf Customized-Trigger} ($n^*_{tri}=5$)} & \multicolumn{7}{c}{{\bf Definable-Trigger}} \\
    %\cline{3-12}
    {\bf Datasets}  &  \multicolumn{1}{c}{$\rho$} & \multicolumn{3}{c|}{{\bf Opt-GDBA}} & \multicolumn{1}{c}{$n_{tri}$} & \multicolumn{2}{c}{{\bf Opt-GDBA}} & \multicolumn{2}{c}{{\bf Rand-GDBA}} & \multicolumn{2}{c}{\bf Rand-GCBA} \\
    %\cline{3-5} \cline{7-12}
   {\bf (No attack)} &  & (MA\ /\ BA) & $n_{tri}$ & $e_{tri}$ & & (MA\ /\ BA) & $e_{tri}$ & (MA\ /\ BA) & $e_{tri}$ & (MA\ /\ BA) & $e_{tri}$ \\ % 
    \hline
    
    & &  & & & 3 & 0.73\ /\ 0.97 (\textcolor{red}{$\uparrow$0.48}) & 1.38 & 0.71\ /\ 0.49 & 2 & 0.71\ /\ 0.38 & 3\\
    % \cline{ 6-12}
    & 10\% & 0.71\ /\ 0.98 (\textcolor{red}{$\uparrow$0.43}) & 4.29 & 3.88 & 4 & 0.72\ /\ 0.98 (\textcolor{red}{$\uparrow$0.43}) & 2.88 & 0.71\ /\ 0.55 & 4 & 0.71\ /\ 0.47 & 6\\
    % \cline{ 6-12}
    & &  & & & 5 & 0.73\ /\ 0.98 (\textcolor{red}{$\uparrow$0.34}) & 4.95 & 0.71\ /\ 0.64 & 6 & 0.73\ /\ 0.51 & 10\\
    \cline{ 2-12}
    & &  & & & 3 & 0.73\ /\ 0.99 (\textcolor{red}{$\uparrow$0.44}) & 1.53 & 0.72\ /\ 0.55 & 2 & 0.72\ /\ 0.53 & 3\\
    % \cline{ 6-12}
    BITCOIN & 20\% & 0.72\ /\ 0.99 (\textcolor{red}{$\uparrow$0.36})& 4.29 & 4.20 & 4 & 0.72\ /\ 0.99 (\textcolor{red}{$\uparrow$0.36}) & 3.08 & 0.71\ /\ 0.63 & 4 & 0.72\ /\ 0.57 & 6\\
    % \cline{ 6-12}
    (MA$=$0.73) & &  & & & 5 & 0.72\ /\ 0.99 (\textcolor{red}{$\uparrow$0.31}) & 5.09 & 0.71\ /\ 0.68 & 6 & 0.71\ /\ 0.61 & 10\\
    \cline{ 2-12}
    & &  & & & 3 & 0.72\ /\ 0.99 (\textcolor{red}{$\uparrow$0.22}) & 1.70 & 0.72\ /\ 0.77 & 2 & 0.72\ /\ 0.66 & 3\\
    % \cline{ 6-12}
    & 30\%  & 0.72\ /\ 0.98 (\textcolor{red}{$\uparrow$0.24})& 4.16 & 4.35& 4 & 0.72\ /\ 1.00 (\textcolor{red}{$\uparrow$0.26}) & 3.27 & 0.71\ /\ 0.74 & 4 & 0.71\ /\ 0.62 & 6\\
    % \cline{ 6-12}
    & &  & & & 5 & 0.72\ /\ 1.00 (\textcolor{red}{$\uparrow$0.15}) & 5.35 & 0.71\ /\ 0.85 & 6 & 0.71\ /\ 0.77 & 10\\
    \hline
    %-----------------------------------------------
    &  &  & & & 3 & 0.71\ /\ 0.83 (\textcolor{red}{$\uparrow$0.45}) & 1.67 & 0.71\ /\ 0.38  & 2 & 0.71\ /\ 0.31 & 3 \\
    % \cline{ 6-12}
    & 10\% & 0.75\ /\ 0.86 (\textcolor{red}{$\uparrow$0.43})& 3.64 & 3.52 & 4 & 0.73\ /\ 0.84 (\textcolor{red}{$\uparrow$0.41}) & 3.33 & 0.73\ /\ 0.43 & 4 & 0.71\ /\ 0.41 & 6\\
    % \cline{ 6-12}
    & &  & & & 5 & 0.71\ /\ 0.88 (\textcolor{red}{$\uparrow$0.41}) & 5.83 & 0.71\ /\ 0.47 & 6 & 0.71\ /\ 0.42 & 10\\
    \cline{ 2-12}
    & &  & & & 3 & 0.74\ /\ 0.85 (\textcolor{red}{$\uparrow$0.37}) & 1.64 & 0.70\ /\ 0.48 & 2 & 0.73\ /\ 0.47 & 3\\
    % \cline{ 6-12}
    MUTAG & 20\% & 0.72\ /\ 0.95 (\textcolor{red}{$\uparrow$0.43})& 3.51 & 2.31 & 4 & 0.71\ /\ 0.85 (\textcolor{red}{$\uparrow$0.33}) & 3.41 & 0.71\ /\ 0.52 & 4 & 0.73\ /\ 0.48 & 6\\
    % \cline{ 6-12}
    (MA$=$0.74) & &  & & & 5 & 0.73\ /\ 0.86 (\textcolor{red}{$\uparrow$0.28}) & 5.88 & 0.73\ /\ 0.58 & 6 & 0.71\ /\ 0.49 & 10\\
    \cline{ 2-12}
    & &  & & & 3 & 0.71\ /\ 0.95 (\textcolor{red}{$\uparrow$0.43}) & 1.60 & 0.71\ /\ 0.52 & 2 & 0.72\ /\ 0.48 & 3\\
    % \cline{ 6-12}
    & 30\% & 0.74\ /\ 0.96 (\textcolor{red}{$\uparrow$0.29})& 3.79 & 2.32 & 4 & 0.71\ /\ 0.97 (\textcolor{red}{$\uparrow$0.30}) & 3.27 & 0.71\ /\ 0.67 & 4 & 0.71\ /\ 0.56 & 6\\
    % \cline{ 6-12}
    & &  & & & 5 & 0.74\ /\ 0.97 (\textcolor{red}{$\uparrow$0.25}) & 5.63 & 0.73\ /\ 0.72 & 6 & 0.73\ /\ 0.63 & 10\\
    \hline
    %---------------------------------------------------------
    
    & &  & & & 3 & 0.73\ /\ 0.82 (\textcolor{red}{$\uparrow$0.48}) & 1.18  & 0.72\ /\ 0.34  & 2 & 0.71\ /\ 0.19  & 3\\
    % \cline{ 6-12}
    & 10\% & 0.74\ /\ 0.89 (\textcolor{red}{$\uparrow$0.54}) & 3.24  &  2.68 & 4 & 0.72\ /\ 0.88 (\textcolor{red}{$\uparrow$0.53}) & 2.15  & 0.71\ /\ 0.35  & 4 &  0.71\ /\ 0.23 & 6\\
    % \cline{ 6-12}
    & &  & & & 5 & 0.73\ /\ 0.91 (\textcolor{red}{$\uparrow$0.56}) & 3.56  & 0.71\ /\ 0.35  & 6 & 0.71\ /\ 0.25  & 10\\
    \cline{ 2-12}
    & &  & & & 3 & 0.71\ /\ 0.87 (\textcolor{red}{$\uparrow$0.40}) &  1.12 & 0.71\ /\ 0.47  & 2 & 0.73\ /\ 0.32  & 3\\
    % % \cline{ 6-12}
    PROTEINS & 20\% & 0.72\ /\ 0.90 (\textcolor{red}{$\uparrow$0.39}) & 3.75  &  2.56 & 4 & 0.72\ /\ 0.90 (\textcolor{red}{$\uparrow$0.39}) & 2.07  &  0.70\ /\ 0.51 & 4 & 0.71\ /\ 0.33  & 6\\
    % \cline{ 6-12}
    (MA$=$0.73) & &  & & & 5 &  0.72\ /\ 0.92 (\textcolor{red}{$\uparrow$0.33}) & 3.36 &  0.71\ /\ 0.59 & 6 & 0.71\ /\ 0.39  & 10\\
    \cline{ 2-12}
    & &  & & & 3 & 0.72\ /\ 0.94 (\textcolor{red}{$\uparrow$0.53}) &  1.05 & 0.71\ /\ 0.41 & 2 & 0.71\ /\ 0.33  & 3\\
    % \cline{ 6-12}
    & 30\% &  0.72\ /\ 0.94 (\textcolor{red}{$\uparrow$0.41})& 3.17  & 2.54  & 4 & 0.71\ /\ 0.94 (\textcolor{red}{$\uparrow$0.41}) & 2.07  &  0.74\ /\ 0.53 & 4 & 0.72\ /\ 0.36  & 6\\
    % \cline{ 6-12}
    & &  & & & 5 & 0.71\ /\ 0.96 (\textcolor{red}{$\uparrow$0.37}) &  3.44 & 0.71\ /\ 0.59 & 6 & 0.72\ /\ 0.37  & 10\\
    \hline
    
    %---------------------------------------------------------
    & &  & & & 3 &  0.72\ /\ 0.78 (\textcolor{red}{$\uparrow$0.50}) &  1.29  & 0.72\ /\ 0.28  & 2 &  0.72\ /\ 0.18 & 3\\
    % \cline{ 6-12}
    & 10\% & 0.72\ /\ 0.78 (\textcolor{red}{$\uparrow$0.45}) & 3.10  &  1.64 & 4 & 0.71\ /\ 0.77 (\textcolor{red}{$\uparrow$0.44}) & 3.12  &  0.71\ /\ 0.33 & 4 & 0.72\ /\ 0.27  & 6\\
    % \cline{ 6-12}
    & &  & & & 5 & 0.72\ /\ 0.80 (\textcolor{red}{$\uparrow$0.47}) & 5.42  &  0.72\ /\ 0.33 & 6 & 0.71\ /\ 0.31  & 10\\
    \cline{ 2-12}
    & &  & & & 3 & 0.73\ /\ 0.76 (\textcolor{red}{$\uparrow$0.38}) & 1.40 &  0.72\ /\ 0.38 & 2 &  0.71\ /\ 0.32 & 3\\
    % \cline{ 6-12}
    DD & 20\% & 0.72\ /\ 0.86 (\textcolor{red}{$\uparrow$0.46}) &  3.19 & 1.57  & 4 &  0.72\ /\ 0.78 (\textcolor{red}{$\uparrow$0.38}) & 2.97  & 0.72\ /\ 0.40  & 4 &  0.72\ /\ 0.33 & 6\\
    % \cline{ 6-12}
    (MA$=0.73$)& &  & & & 5 & 0.73\ /\ 0.87 (\textcolor{red}{$\uparrow$0.43}) & 5.04 &  0.73\ /\ 0.44 & 6 & 0.72\ /\ 0.36  & 10\\
    \cline{ 2-12}
    & &  & & & 3 & 0.72\ /\ 0.83 (\textcolor{red}{$\uparrow$0.35}) &  1.51 & 0.72\ /\ 0.48 & 2 &  0.71\ /\ 0.41 & 3\\
    % \cline{ 6-12}
    & 30\% &  0.73\ /\ 0.90 (\textcolor{red}{$\uparrow$0.41})&  3.57 & 1.79  & 4 & 0.72\ /\ 0.90 (\textcolor{red}{$\uparrow$0.41}) &  3.26 & 0.71\ /\ 0.49  & 4 & 0.72\ /\ 0.44 & 6\\
    % \cline{ 6-12}
    & &  & & & 5 &  0.72\ /\ 0.92 (\textcolor{red}{$\uparrow$0.36}) &  5.22 & 0.72\ /\ 0.56 & 6 &  0.72\ /\ 0.48  & 10\\
    \hline
    %---------------------------------------------------------
    & &  & & & 3 & 0.74\ /\ 0.80 (\textcolor{red}{$\uparrow$0.36}) & 1.71 & 0.72\ /\ 0.44 & 2 & 0.73\ /\ 0.37 & 3\\
    % \cline{ 6-12}
    & 10\% & 0.72\ /\ 0.83 (\textcolor{red}{$\uparrow$0.30})& 4.43 & 4.50 & 4 & 0.74\ /\ 0.80 (\textcolor{red}{$\uparrow$0.27}) & 3.28 & 0.73\ /\ 0.53 & 4 & 0.71\ /\ 0.34 & 6\\
    % \cline{ 6-12}
    & &  & & & 5 & 0.75\ /\ 0.82 (\textcolor{red}{$\uparrow$0.23}) & 5.85 & 0.73\ /\ 0.59 & 6 & 0.73\ /\ 0.49 & 10\\
    \cline{ 2-12}
    & &  & & & 3 & 0.74\ /\ 0.81 (\textcolor{red}{$\uparrow$0.30}) & 1.77 & 0.71\ /\ 0.51 & 2 & 0.73\ /\ 0.37 & 3\\
    % \cline{ 6-12}
    COLLAB & 20\% & 0.73\ /\ 0.86 (\textcolor{red}{$\uparrow$0.32})& 4.68 & 4.51 & 4 & 0.73\ /\ 0.84 (\textcolor{red}{$\uparrow$0.30}) & 3.34 & 0.73\ /\ 0.54 & 4 & 0.71\ /\ 0.37 & 6\\
    % \cline{ 6-12}
    (MA$=$0.75) & &  & & & 5 & 0.73\ /\ 0.85 (\textcolor{red}{$\uparrow$0.29})& 5.72 & 0.73\ /\ 0.56 & 6 & 0.71\ /\ 0.42 & 10\\
    \cline{ 2-12}
    & &  & & & 3 & 0.71\ /\ 0.85 (\textcolor{red}{$\uparrow$0.25}) & 1.73 & 0.69\ /\ 0.60 & 2 & 0.73\ /\ 0.41 & 3\\
    % \cline{ 6-12}
    & 30\% & 0.71\ /\ 0.92 (\textcolor{red}{$\uparrow$0.31})& 4.59 & 4.72 & 4 & 0.71\ /\ 0.90 (\textcolor{red}{$\uparrow$0.29}) & 3.58 & 0.72\ /\ 0.61 & 4 & 0.71\ /\ 0.46 & 6\\
    % \cline{ 6-12}
    & &  & & & 5 & 0.71\ /\ 0.91 (\textcolor{red}{$\uparrow$0.26}) & 5.99 & 0.71\ /\ 0.65& 6 & 0.71\ /\ 0.47 & 10\\
    \hline
    %---------------------------------------------------------
    & &  & & & 3 & 0.52\ /\ 0.85 (\textcolor{red}{$\uparrow$0.42})  & 1.51 & 0.52\ /\ 0.43  & 2 & 0.52\ /\ 0.26  & 3\\
    % \cline{ 6-12}
    & 10\% & 0.51\ /\ 0.89 (\textcolor{red}{$\uparrow$0.38}) & 4.18  & 3.42  & 4 & 0.52\ /\ 0.90 (\textcolor{red}{$\uparrow$0.39}) & 3.41  & 0.51\ /\ 0.51  & 4 & 0.52\ /\ 0.29  & 6\\
    % \cline{ 6-12}
    & &  & & & 5 & 0.53\ /\ 0.90 (\textcolor{red}{$\uparrow$0.38}) & 5.39 & 0.52\ /\ 0.52  & 6 &  0.53\ /\ 0.36 & 10\\
    \cline{ 2-12}
    & &  & & & 3 & 0.52\ /\ 0.87 (\textcolor{red}{$\uparrow$0.35}) & 1.62 & 0.52\ /\ 0.52  & 2 &  0.52\ /\ 0.37 & 3\\
    % \cline{ 6-12}
    RDT-M5K &20\% & 0.52\ /\ 0.90 (\textcolor{red}{$\uparrow$0.33})& 4.51  & 3.59  & 4 & 0.52\ /\ 0.89 (\textcolor{red}{$\uparrow$0.32}) & 3.27  & 0.52\ /\ 0.57 & 4 & 0.52\ /\ 0.40  & 6\\
    % \cline{ 6-12}
    (MA$=0.53$)& &  & & & 5 &  0.52\ /\ 0.90 (\textcolor{red}{$\uparrow$0.33}) & 5.45 & 0.53\ /\ 0.57  & 6 & 0.52\ /\ 0.42  & 10\\
    \cline{ 2-12}
    & &  & & & 3 &  0.53\ /\ 0.89 (\textcolor{red}{$\uparrow$0.22}) & 1.64 & 0.52\ /\ 0.67 & 2 & 0.52\ /\ 0.45  & 3\\
    % \cline{ 6-12}
    & 30\% &  0.52\ /\ 0.92 (\textcolor{red}{$\uparrow$0.25})& 4.49  & 3.86 & 4 & 0.52\ /\ 0.90 (\textcolor{red}{$\uparrow$0.23}) &  3.33 &  0.51\ /\ 0.67 & 4 & 0.51\ /\ 0.47  & 6\\
    % \cline{ 6-12}
    & &  & & & 5 &  0.52\ /\ 0.92 (\textcolor{red}{$\uparrow$0.14}) & 5.59  & 0.51\ /\ 0.78 & 6 &  0.51\ /\ 0.55 & 10\\
    \bottomrule
    %---------------------------------------------------------
    \end{tabular}
    \label{table:GraphDBA-results-full}  
   \vspace{-4mm}
\end{table*}

\begin{table*}[!t]
\small
%\footnotesize
\renewcommand{\arraystretch}{1}
%\addtolength{\tabcolsep}{-2pt}
\caption{We compare the raw way~\cite{xu2022more} and our way of backdoored testing in Rand-GDBA. 
}
\vspace{-2mm}
\centering
    \begin{tabular}{c||c|c|c|c|c|c|c}
    \toprule
    Attack &Dataset &BITCOIN  &  MUTAG& PROTEINS & DD & COLLAB & RDT-M5K \\ 
    \hline
    \hline
    &MA/BA &  0.72/0.85 & 0.71/0.61  & 0.71/0.69 &0.72/0.58 & 0.73/0.68  &  0.52/0.69 \\
    %\cline{2-8}
    Rand-GDBA (raw)~\cite{xu2022more} &NetSim ($\uparrow$) &  0.32 &  0.41  & 0.55  &  0.95 & 0.87  & 0.90  \\
    %\cline{2-8}
    &DeltaCon ($\uparrow$) &  0.44 &  0.53  &  0.69 & 0.96  & 0.93& 0.94 \\
    \hline
    &MA/BA &  0.71/0.63  & 0.71/0.52  &  0.70/0.51 & 0.72/0.40 &  0.73/0.54  & 0.52/0.57   \\
    % \cline{2-8}
    Rand-GDBA (ours) &NetSim ($\uparrow$) & 0.52 &  0.88 &  0.86  &  0.99  &  0.99  &  0.98  \\
    % \cline{2-8}
    &DeltaCon ($\uparrow$) &  0.63  &  0.90  &  0.91  &  0.99  & 0.99 & 0.98  \\
    \hline \hline
    &MA/BA &  0.72/0.99  & 0.71/0.85  &  0.72/0.90 & 0.72/0.78 &  0.73/0.84  & 0.52/0.89   \\
    % \cline{2-8}
     Our Opt-GDBA  & NetSim ($\uparrow$) & 0.55 &  0.90 &  0.88  &  0.99  &  0.99  &  0.98  \\
    % \cline{2-8}
    (Definable-Trigger) &DeltaCon ($\uparrow$) &  0.65  &  0.93  &  0.94  &  0.99  & 0.99 & 0.99  \\
    \hline 
    &MA/BA &  0.72/0.99  & 0.72/0.95  &  0.72/0.90 & 0.72/0.86 &  0.73/0.86  & 0.52/0.90   \\
    % \cline{2-8}
     Our Opt-GDBA  & NetSim ($\uparrow$) & 0.54 &  0.87 &  0.86  &  0.99  &  0.99  &  0.98  \\
    % \cline{2-8}
    (Customized-Trigger) & DeltaCon ($\uparrow$) &  0.64  &  0.92  &  0.91  &  1.00  & 0.99 & 0.98  \\
    \bottomrule
    \end{tabular}
\label{table:comparsion experiment}  
\vspace{-2mm}
\end{table*}

\begin{table}[h]
\small
\renewcommand{\arraystretch}{0.9}
\addtolength{\tabcolsep}{-2pt}
\caption{Results of Opt-GDBA with less \#malicious clients (4 and 1) and less poisoning rate (10\%). 
}
\vspace{-2mm}
\centering
    \begin{tabular}{c||c c c|c c c}
    \toprule  
    {\bf Dataset} & \multicolumn{3}{c}{Customized-Trigger} & \multicolumn{3}{c}{Definable-Trigger} \\
    % \cline{2-7}
    {\bf (4 mali. clients)}& (MA\ /\ BA) & $n_{tri}$ & $e_{tri}$ & (MA\ /\ BA) & $n_{tri}$ & $e_{tri}$ \\ 
    \hline
    BITCOIN & 0.72\ /\ 0.90 & 4.29 & 3.88 & 0.72\ /\ 0.89 & 4 & 2.20 \\
    % \hline
    MUTAG  & 0.73\ /\ 0.85 & 3.41 & 2.10 & 0.71\ /\ 0.82 & 4 & 1.53 \\
    % \hline
    PROTEINS  & 0.72\ /\ 0.81 & 3.96 & 2.72 & 0.73\ /\ 0.86 & 4 & 2.33 \\
    % \hline    
    DD  & 0.72\ /\ 0.74 & 3.04 & 1.66 & 0.71\ /\ 0.77 & 4 & 3.14 \\
    % \hline
     COLLAB  &  0.72\ /\ 0.77 & 4.52  & 4.67  &  0.72\ /\ 0.76  & 4 & 3.61  \\
    % \hline    
    RDT-M5K  & 0.52\ /\ 0.84 & 4.32 & 3.78  & 0.53\ /\ 0.82  & 4 &  3.42 \\
    \hline
    \hline
    {\bf (1 mali. client)} & (MA\ /\ BA) & $n_{tri}$ & $e_{tri}$ & (MA\ /\ BA) & $n_{tri}$ & $e_{tri}$ \\ 
    \hline
    BITCOIN & 0.72\ /\ 0.74 & 4.92 & 2.61 & 0.72\ /\ 0.70 & 4 & 1.53 \\
    % \hline
    MUTAG  & 0.72\ /\ 0.76 & 3.39 & 2.37 & 0.73\ /\ 0.73 & 4 & 3.18\\
    % \hline
    PROTEINS  & 0.72\ /\ 0.78  & 3.88 & 2.26 & 0.72\ /\ 0.70 & 4 & 2.05 \\
    % \hline
    DD  & 0.72\ /\ 0.67 & 3.37 & 1.75 & 0.72\ /\ 0.69 & 4 & 3.25 \\  
    % \hline
    COLLAB  &  0.73\ /\ 0.70  &  4.34  & 4.47  & 0.72\ /\ 0.68 & 4 &  3.44 \\
    % \hline
    RDT-M5K  & 0.52\ /\ 0.72 & 4.57  & 3.65  & 0.52\ /\ 0.71 & 4 & 3.50  \\  
    \bottomrule
    \end{tabular}
    \label{table: less clients}
   \vspace{-4mm}
\end{table}

\begin{table}[h]
\small
\renewcommand{\arraystretch}{0.9}
\addtolength{\tabcolsep}{-1pt}
\caption{Attack results of Opt-GDBA with varying numbers of clients: $C=20$ and $60$ ($30$ for MUTAG due to its limited size). The default value is $C=40$ in the paper.} 
\vspace{-2mm}
\centering
    \begin{tabular}{c||c|c c c|c c c }
    \toprule 
    {\bf Dataset} & \multicolumn{1}{c}{$C$} & \multicolumn{3}{c}{Customized-Trigger} & \multicolumn{3}{c}{Definable-Trigger} \\
    % \cline{2-7} 
    & & (MA\ /\ BA) & $n_{tri}$ & $e_{tri}$ & (MA\ /\ BA) & $n_{tri}$ & $e_{tri}$ \\ 
    \hline
    BITCOIN & $20$ & 0.71\ /\ 0.95 & 3.99 & 2.20 & 0.72\ /\ 0.94 & 4 & 1.72 \\    
    & $60$ & 0.72\ /\ 0.93  & 4.19  &  2.57 &  0.72\ /\ 0.95 &  4 & 2.64  \\
    % \hline
    \hline
    MUTAG & $20$ & 0.71\ /\ 0.87  & 3.43 & 2.92  & 0.71\ /\ 0.85  & 4   & 3.96  \\    
    & $30$ &  0.74\ /\ 0.90 &  3.50  &  2.81  &  0.74\ /\ 0.88  & 4  & 4.07   \\
    \hline
    PROTEINS & $20$ & 0.72\ /\ 0.90 & 3.87  & 2.28  & 0.72\ /\ 0.89  & 4  & 2.26  \\    
    & $60$ &  0.72\ /\ 0.93 &  3.23  &  2.47  &  0.72\ /\ 0.91 &  4 & 2.14   \\
    % \hline
    \hline
    DD & $20$ & 0.72\ /\ 0.76 & 3.04   & 1.67  & 0.72\ /\ 0.77 & 4   & 3.03   \\    
    % \hline
    & $60$ &  0.72\ /\ 0.81 & 3.42   & 1.51    & 0.72\ /\ 0.78 & 4   & 2.97    \\
    \hline %\hline 
     COLLAB & $20$ &  0.72\ /\ 0.86 &  4.66   &  4.47  &   0.72\ /\ 0.87 & 4  &  3.47  \\    
    & $60$ &  0.73\ /\ 0.85  &   4.53  &   4.62  &  0.72\ /\ 0.85  &  4 &  3.61   \\
    % \hline
    \hline
    RDT-M5K & $20$ & 0.53\ /\ 0.91  &  4.70  &  3.81  &   0.52\ /\ 0.90 & 4   & 3.33    \\    
    % \hline
    & $60$ &  0.52\ /\ 0.90  &  4.44   &  3.67 & 0.52\ /\ 0.90  & 4   & 3.92     \\
    \bottomrule
    \end{tabular}
    \label{table: varying numbers of clients}
   %\vspace{-2mm}
\end{table}

\begin{table}[!h]
\renewcommand{\arraystretch}{0.9}
\small
\addtolength{\tabcolsep}{-2.5 pt}
\caption{Attack results of Opt-GDBA on non-IID and imbalanced datasets. 
}
\vspace{-2mm}
\centering
    \begin{tabular}{c||c|c c c|c c c}
    \toprule 
    {\bf Dataset} & \multicolumn{1}{c}{Scenes} & \multicolumn{3}{c}{Customized-Trigger} & \multicolumn{3}{c}{Definable-Trigger} \\
    %\cline{2-7}
    & & (MA\ /\ BA) & $n_{tri}$ & $e_{tri}$ & (MA\ /\ BA) & $n_{tri}$ & $e_{tri}$ \\ 
    \hline
    BITCOIN & {  Non-IID}  & 0.72\ /\ 0.98  & 3.55  & 2.60  & 0.72\ /\ 0.96  & 4 & 2.75 \\    
    & {  Imbalanced}  & 0.72\ /\ 0.94  & 4.10  &  3.34 &  0.72\ /\ 0.97 & 4  & 3.44  \\
    \hline
    MUTAG & {  Non-IID} &  0.71\ /\ 0.95 & 3.51  & 2.81  & 0.70\ /\ 0.94 & 4  &  3.46 \\    
    &  {  Imbalanced}   &  0.74\ /\ 0.93  & 3.73   & 3.28   &  0.74\ /\ 0.90  & 4  & 2.51   \\
    \hline
    PROTEINS & {  Non-IID}  &  0.72\ /\ 0.94 & 3.77  & 2.66  & 0.72\ /\ 0.93 & 4  & 2.68  \\    
    & {  Imbalanced}   &  0.72\ /\ 0.90  &  3.29  &  1.96  &  0.72\ /\ 0.89  & 4  & 2.45   \\
    \hline
    DD &{  Non-IID}  &  0.72\ /\ 0.87 &  3.12  &  1.38  & 0.72\ /\ 0.75 & 4 & 2.53  \\    
    %\hline
    & {  Imbalanced}   &  0.72\ /\ 0.82 &  3.24   & 1.27   &   0.72\ /\ 0.74 &  4 & 2.48    \\
    \hline
    COLLAB &{  Non-IID}  &  0.72\ /\ 0.85 &  4.17  & 4.42  & 0.73\ /\ 0.84 & 4  &  3.76 \\    
    %\hline
    & {  Imbalanced}   &  0.72\ /\ 0.83  &  4.25  &  4.63  & 0.72\ /\ 0.82  &  4  &  3.37  \\
    \hline
    RDT-M5K &{  Non-IID}  &  0.52\ /\ 0.89 & 4.07 &  3.53  & 0.52\ /\ 0.88 & 4  &  3.61  \\    
    %\hline
    & {  Imbalanced}   &  0.52\ /\ 0.90 &  4.33  &  3.67  &  0.52\ /\ 0.90 &  4  & 3.44 \\
    \bottomrule
    \end{tabular}
    \label{table: Non-IID and Imbalanced}
    %\vspace{-1mm}
\end{table}

%\vspace{-4mm}
\section{More Experimental Results}
\label{app:results}

\subsection{More Attack Results}
\label{app:attackresults}

\noindent {\bf Comprehensive attack results:} {\bf Table~\ref{table:GraphDBA-results-full}} shows the comprehensive results of the compared attacks in all experimental settings. 
Overall, all attacks can achieve a close MA as the MA under no attack (their difference is $\leq 4$ in all cases). 
Rand-GDBA outperforms Rand-GCBA with a gain from $2\%$ to $29\%$, indicating the distributed graph backdoor is more effective than the centralized counterpart.  
Further, our Opt-GDBA  achieves much better BA and is more stealthy than Rand-GDBA---our attack's gain is ranging from $14\%$ to $56\%$ with an average more than 30\%. 
Comparing the two proposed trigger location learning schemes in Opt-GDBA,
the Customized-Trigger is more stealthy than the Definable-Trigger, i.e., its generated triggers has a smaller size under the same MA/BA.  

\noindent {\bf Comparing \cite{xu2022more} and our way for backdoor testing in Rand-GDBA:} {\bf Table~\ref{table:comparsion experiment}} shows the comparison results.  
The original way \cite{xu2022more} combines the local triggers in all malicious clients into a global trigger.  
That is, the total number of edges (nodes) in the global trigger is the sum of that in all the local triggers. Such a large trigger ensures Rand-GDBA (raw) exhibiting a higher BA than Rand-GDBA (ours) for backdoor testing, but also largely changing the structure of testings graphs.   
As a reference, we also show the results of our Opt-GDBA. Still, our Opt-GDBA 
not only achieves much higher BA but is also more stealthy. 

\noindent {\bf Attack results of Opt-GDBA on less \#malicious clients and poison rate:}
In the paper, we follow 
prior works~\cite{xie2019dba,zhang2021backdoor} by setting $\rho=20\% $ (8 malicious clients) and 50\% poison rate. Here, we add results with less  (4 and 1) malicious clients and less (10\%) poison rate  and the backdoor attack results of Opt-GDBA are shown in {\bf Table~\ref{table: less clients}.} 
We can see: 1) Our Opt-GDBA remains \emph{highly effective} with 4 malicious client and 10\% poison rate, i.e., high BA and no/marginal MA loss; 2) A \emph{slight/moderate reduction} in BA is observed with only 1 malicious client. This is understandable as a single malicious client lacks the strong capability to execute a distributed backdoor attack.

\noindent {\bf Attack results on varying number of clients.}
{\bf Table~\ref{table: varying numbers of clients}} shows the Opt-GDBA 
 results on 20 and 60 clients (default value is 40). We can see Opt-GDBA consistently maintains both high MA and BA, showing its performance is insensitive to the number of  clients.

\noindent {\bf Attack results on non-IID and imbalanced datasets.} {\bf Table~\ref{table: Non-IID and Imbalanced}} shows the  attack results of Opt-GDBA on non-IID and imbalanced datasets. The non-IID datasets are generated by letting each client holds a single label data, while the imbalanced datasets are generated via letting different clients hold different ratios of data from different classes. 
We can see Opt-GDBA remains effective with non-IID and imbalanced datasets.

\begin{table*}[!t]
\small
\renewcommand{\arraystretch}{0.9}
\addtolength{\tabcolsep}{-4pt}
\caption{Attack results of Opt-GDBA against model pruning (with the pruning rate $\alpha$) and distillation. 
}
\vspace{-2mm}
\centering
\begin{tabular}{c||c c|c|c c}
\toprule 
{\bf Dataset} & {\bf Customized-Trigger} & \multicolumn{1}{c}{\bf Definable-Trigger} & \multicolumn{1}{c}{\bf Dataset} &  {\bf Customized-Trigger} & {\bf Definable-Trigger}  \\ 
% \cline{2-3} \cline{5-6} 
                  & (MA\ /\ BA) & (MA\ /\ BA) &               & (MA\ /\ BA) &  (MA\ /\ BA)    \\ \hline
    BITCOIN ({\bf no attack}) & 0.72\ /\ 0.90 & 0.72\ /\ 0.89 & DD ({\bf no attack}) &  0.72\ /\ 0.74  &  0.71\ /\ 0.77  \\
     BITCOIN ($\alpha=20\%$) &  0.57\ /\ 0.73  & 0.55\ /\ 0.69  &  DD ($\alpha=20\%$) &   0.52\ /\ 0.62  & 0.48\ /\ 0.67   \\  
    % \hline 
    BITCOIN ($\alpha=15\%$) &  0.67\ /\ 0.85  &  0.69\ /\ 0.88  & DD ($\alpha=15\%$) &  0.55\ /\ 0.71  & 0.52\ /\ 0.69  \\    
    % \hline 
    BITCOIN ($\alpha=10\%$) & 0.69\ /\ 0.90  &  0.69\ /\ 0.89 &  DD ($\alpha=10\%$) &  0.62\ /\ 0.72&  0.64\ /\ 0.76 \\    
    BITCOIN {\bf (distillation)}  & 0.72\ /\ 0.87 &  0.72\ /\ 0.89 & DD {\bf (distillation)} &  0.72\ /\ 0.72 & 0.72\ /\ 0.75 \\
    \hline 
    % \hline
    MUTAG ({\bf no attack}) & 0.73\ /\ 0.85  & 0.71\ /\ 0.82 & COLLAB ({\bf no attack}) & 0.72\ /\ 0.80 & 0.73\ /\ 0.79 \\
    MUTAG ($\alpha=20\%$) & 0.33\ /\ 0.00 &  0.33\ /\ 0.00 & COLLAB ($\alpha=20\%$) & 0.57\ /\ 0.65 & 0.59\ /\ 0.71 \\  
    % \hline
    MUTAG ($\alpha=15\%$) &  0.65\ /\ 0.65   &  0.66\ /\ 0.80 & COLLAB ($\alpha=15\%$)  & 0.63\ /\ 0.75  & 0.65\ /\ 0.74  \\    
    % \hline
    MUTAG ($\alpha=10\%$) & 0.72\ /\ 0.82   & 0.68\ /\ 0.82 & COLLAB ($\alpha=10\%$)  & 0.71\ /\ 0.80  &  0.72\ /\ 0.78  \\    
    MUTAG {\bf (distillation)}  &  0.73\ /\ 0.79 &  0.73\ /\ 0.80 & COLLAB {\bf (distillation)}  &  0.72\ /\ 0.79 & 0.72\ /\ 0.78 \\
    \hline
    PROTEINS ({\bf no attack}) &  0.72\ /\ 0.81  & 0.73\ /\ 0.86  & RDT-M5K ({\bf no attack}) &  0.52\ /\ 0.84 & 0.53\ /\ 0.82 \\
    PROTEINS ($\alpha=20\%$) &  0.58\ /\ 0.54   & 0.59\ /\ 0.65 & RDT-M5K ($\alpha=20\%$)  &  0.34\ /\ 0.67 &  0.35\ /\ 0.62    \\  
    % \hline
    PROTEINS ($\alpha=15\%$) &  0.61\ /\ 0.65  &  0.62\ /\ 0.77 & RDT-M5K ($\alpha=15\%$)  & 0.41\ /\ 0.74  &  0.43\ /\ 0.72  \\    
    % \hline
    PROTEINS ($\alpha=10\%$) & 0.65\ /\ 0.72  & 0.71\ /\ 0.86 & RDT-M5K ($\alpha=10\%$)  & 0.52\ /\ 0.81  & 0.51\ /\ 0.79    \\    
    PROTEINS {\bf (distillation)} &  0.72\ /\ 0.80 & 0.74\ /\ 0.86 & RDT-M5K {\bf (distillation)}  &  0.52\ /\ 0.81 & 0.52\ /\ 0.80   \\
    \bottomrule
    \end{tabular}
    \label{table: empirical defenses}
   %\vspace{-2mm}
\end{table*}

\noindent {\bf Defending Opt-GDBA with model pruning and distillation.} {\bf Table~\ref{table: empirical defenses}} shows the results. We can see: 1) model pruning is ineffective against Opt-GDBA with a small pruning rate $\alpha$. With a larger pruning rate (e.g., $\alpha=20\%$), though BA is somewhat reduced, MA is significantly reduced; and  
2) distillation \cite{li2020neural} is also ineffective. 
% \vspace{-5mm}

%==============================================
\begin{figure*}[!t]
\centering	
\includegraphics[width=0.85\textwidth]{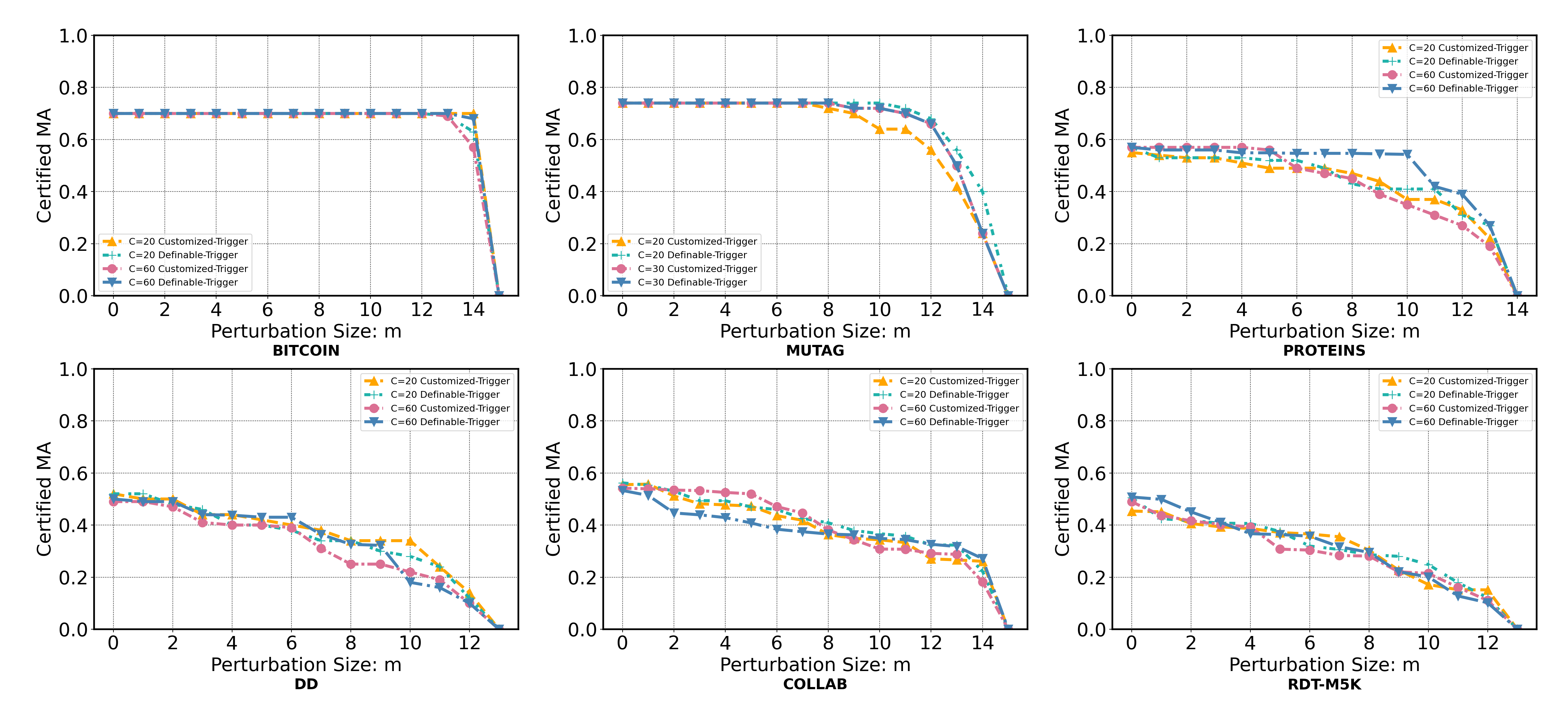}
\vspace{-2mm}
\caption{Certified MA vs. varying number of clients.
}
\label{fig:plot-defense-client-n}
\vspace{-2mm}
\end{figure*}

\begin{figure*}[h]
\centering	
\includegraphics[width=0.85\textwidth]{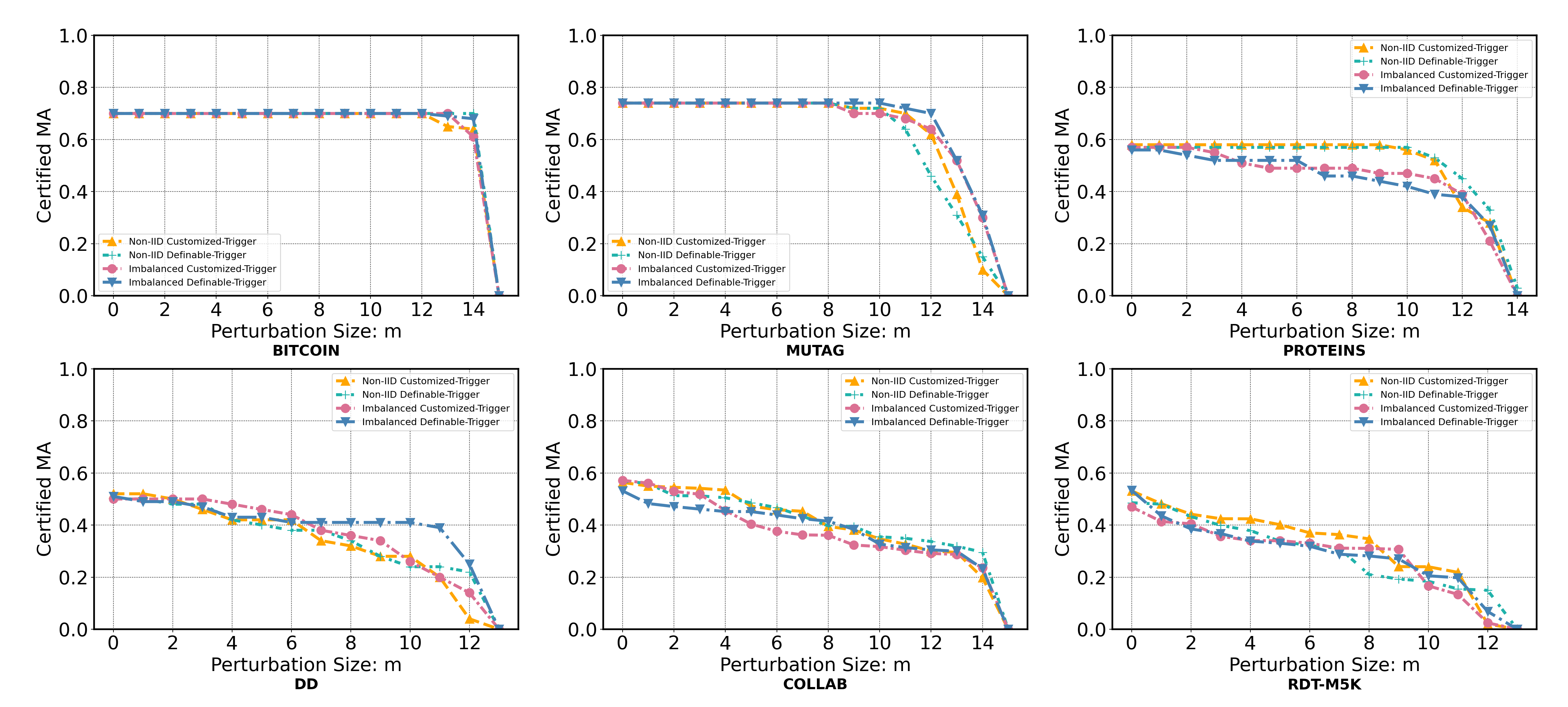}
\vspace{-2mm}
\caption{Certified MA vs. Non-IID \& Unbalanced datasets.
}
\label{fig:plot-defense-non-IID}
%\vspace{-2mm}
\end{figure*}

\subsection{More Defense Results}
\label{app:defenseresults}

\noindent {\bf Defense results on varying number of clients.} {\bf Figure~\ref{fig:plot-defense-client-n}}
shows our defense results vs varying number of clients $C$.  
We can see our defense is effective and relatively stable across different $C$. 

\noindent {\bf Defense results on non-IID on imbalanced datasets.}
{\bf Figure~\ref{fig:plot-defense-non-IID}} shows our defense results on non-IID on imbalanced datasets. The non-IID datasets are generated by letting each client holds a single label data, while the imbalanced datasets are generated via letting different clients hold different ratios of data from different classes (e.g., for binary-class data, client i holds 90\% label A data, while client 2 holds 90\% label B data).
Still, our defense is  effective. 

\end{document}